\documentclass[prd,aps,twocolumn,a4paper,showkeys,nofootinbib]{revtex4-2}

\usepackage{graphicx,psfrag}
\usepackage{mathrsfs}
\usepackage{amsmath,amsfonts,amssymb}
\usepackage{multirow}
\usepackage{comment}
\usepackage{subfigure}
\usepackage{appendix}
\usepackage{hyperref}
\usepackage{ragged2e}

\usepackage{enumitem}

\newcommand{\be}{\begin{equation}}
\newcommand{\ee}{\end{equation}}
\newcommand{\bea}{\begin{eqnarray}}
\newcommand{\eea}{\end{eqnarray}}
\newcommand{\bel}{\begin{align}}
\newcommand{\eel}{\end{align}}

\def\l{\ell}
\def\lm{\ell m}

\def\GMc2{{\rm G M_{\odot} c^{-2}}}

\def\kt2{\kappa^\text{T}_2}

\usepackage{pifont} 

\newcommand{\BAM}[1]{\texttt{BAM:{#1}}} 

\newcommand{\EFL}[1]{EFL{#1}} 
\usepackage{color}
\definecolor{cyan}{rgb}{0,0.9,0.9}
\definecolor{orange}{rgb}{0.9,0.5,0}
\definecolor{magenta}{rgb}{1,0,1}
\definecolor{purple}{rgb}{0.8,0.4,0.8}
\definecolor{gray}{rgb}{0.8242,0.8242,0.8242}

\begin{document}

\title{Eccentricity reduction of binary neutron star initial data with the entropy based flux limiting scheme}

\author{Georgios \surname{Doulis}$^{1}$}
\author{Sebastiano \surname{Bernuzzi}$^{2}$}
\author{Wolfgang \surname{Tichy}$^{3}$}

\affiliation{${}^1$Institut f{\"u}r Theoretische Physik, Goethe-Universit{\"a}t Frankfurt, 60438 Frankfurt am Main}
\affiliation{${}^2$Theoretisch-Physikalisches Institut, Friedrich-Schiller-Universit{\"a}t Jena, 07743 Jena}
\affiliation{${}^3$Department of Physics, Florida Atlantic University, Boca Raton, FL 33431}

\date{\today}

\begin{abstract}
The construction of high-resolution shock-capturing schemes is vital in producing highly accurate gravitational waveforms from neutron star binaries. The entropy based flux limiting (EFL) scheme is able to perform fast converging binary neutron star merger simulations reaching up to fourth-order convergence in the gravitational waveform phase. In these results the EFL method was used only in the dynamical evolution of initial data constructed with the Lorene library. Here, we extend the use of the EFL method to the construction of eccentricity reduced initial data for neutron star binaries and present several new BNS simulations resulting from such initial data and show for the first time up to optimal fifth-order convergence in the gravitational waveform phase.
\end{abstract}


\maketitle

\section{Introduction}

Binary neutron star (BNS) coalescences are one of the confirmed \cite{TheLIGOScientific:2017qsa,GBM:2017lvd} sources of gravitational waves (GWs) detected by the LIGO-VIRGO interferometers. Numerical relativity (NR) is able to produce merger waveforms that replicate these GWs accurately enough for BNS parameter estimation. The production of highly accurate numerical waveforms plays a vital role in understanding the properties of neutron star (NS) binaries during the inspiral, merger and postmerger regime. In this spirit, several high-resolution shock-capturing (HRSC) schemes \cite{Toro:1999} have been developed over the years aiming at increasing the accuracy and the convergence properties of the numerically produced waveforms. Going beyond second-order convergence is currently the major challenge for the existing HRSC schemes, which has been pointed out long ago, but addressed only in a few places \cite{Bernuzzi:2012ci,Radice:2013hxh,Radice:2013xpa,Bernuzzi:2016pie}. Here, we focus on a specific family of flux altering HRSC methods: the flux-limiters \cite{Sweby:1984a}, which have been proven very successful in dealing with shocks and suppressing spurious oscillations.

In \cite{Doulis:2022vkx} a fast converging HRSC scheme that utilises the local entropy production in the design of an entropy based flux-limiter was proposed. The EFL scheme developed from an entropy method that started as an artificial viscosity method \cite{Guermond:2008,Guermond:2011} and grew with time to a flux-limiting scheme \cite{Guercilena:2016fdl}. In \cite{Guercilena:2016fdl}, for the first time, entropy was used in the design of a flux-limiter. The EFL scheme, apart from overcoming the shortcomings of the method developed in \cite{Guercilena:2016fdl}, is considerably more automated and simpler in its implementation. These features make the EFL method universally applicable to different special and general relativistic scenarios. The EFL scheme was successfully used in \cite{Doulis:2022vkx} to evolve several different BNS systems and managed to deliver up to fourth-order convergent waveforms in the dominant $(2,2)$-mode and in the subdominant $(3,2)$- and $(4,4)$-mode.


In the present work we extend the use of the EFL method to the construction of BNS initial data through an eccentricity reduction procedure. In \cite{Doulis:2022vkx}, the EFL scheme was exclusively used for the temporal evolution of BNS initial data computed with the Lorene library \cite{Gourgoulhon:2000nn}. Here, for the first time, we involve the EFL method in the eccentricity reduction procedure of initial data sets with the SGRID code \cite{Tichy:2012rp} and study the effect of this action on the accuracy and the convergence properties of the produced merger waveforms.

The article is structured as follows. Theoretical and numerical aspects of the EFL method are discussed in \autoref{sec:EFL_method}. The way to perform eccentricity reduction in the construction of SGRID initial data is comprehensively detailed in \autoref{sec:ID}. \autoref{sec:BNS} contains our results of the first BNS simulations where the EFL method is not only used for the temporal evolution, but also in the eccentricity reduction of the initial data. Finally, we conclude in \autoref{sec:conclusions}.

Throughout this work we use geometric units. We set $c = G = 1$ and the masses are expressed in terms of solar masses $M_\odot$.

\section{EFL Method}
\label{sec:EFL_method}

We start from the equations of general relativistic hydrodynamics written in conservation form \cite{Banyuls:1997zz}: 
\begin{equation}
 \label{eq:conserv_PDE}
 \partial_t \textbf{Q} + \partial_i \textbf{F}^i(\textbf{Q}) = \textbf{S},
\end{equation}
where the summation is performed over the spatial dimensions and 
\begin{eqnarray*}
 \frac{\textbf{Q}}{\sqrt{\gamma}} &=& \rho\,W\left(1, h\,W u_j, h\,W - 1 -\frac{p}{\rho W}\right)=\left(D,S_j,\tau\right),\\ 
 \frac{\textbf{F}^i}{\sqrt{\gamma}} &=& \left(\mathrm{v}^i D, \mathrm{v}^i S_j + \alpha p \delta^i_j, \mathrm{v}^i \tau + \alpha p u^i\right),\\
 \frac{\textbf{S}}{\sqrt{\gamma}} &=& \alpha \,\left(0, \Gamma^\mu_{\nu j}\,T^\nu_\mu, \alpha\,[T^{0\mu}\,\partial_\mu \ln\alpha - \Gamma^0_{\mu\nu}T^{\mu\nu}]\right)
\end{eqnarray*}  
are the vectors of the conserved variables, the physical fluxes and the sources, respectively. In the above definitions we introduced the rest-mass density $\rho$, the pressure $p$, the 3-velocity $u^i$ of the fluid, the specific enthalpy $h = 1 + \epsilon + p/\rho$ with $\epsilon$ the specific internal energy, the Lorentz factor $W = (1 - u_iu^i)^{1/2}$, the determinant $\gamma$ of the 3-metric $\gamma_{ij}$, the lapse function $\alpha$, the shift vector $\beta^i$, the Kronecker delta $\delta^i_j$, the notation $\mathrm{v}^i=\alpha u^i-\beta^i$, the energy-momentum tensor $T_{\mu\nu}$ and the Christoffel symbols $\Gamma^\lambda_{\mu\nu}$ associated with the metric $g_{\mu\nu}$.

The EFL method is based on the idea of expressing the numerical fluxes resulting from the semi-discretisation of \eqref{eq:conserv_PDE} as a superposition of an unstable and a stable flux. As discussed in \cite{Doulis:2022vkx} the former is approximated by a high-order accurate flux and the latter by a low- or high-order accurate flux. The weight regulating the superposition of the two fluxes is computed from the entropy produced by the conservation law under study. In essence, the entropy of the system is a ``shock detector" that is used as a switch indicating when to move from the unstable high-order scheme to the stable one.

In accordance with \cite{Doulis:2022vkx}, we use the conservative finite-difference formula
\begin{equation}
 \label{eq:spatial_disc}
  \partial_x F^x_i = \frac{\hat f_{i+1/2} - \hat f_{i-1/2}}{h}
\end{equation}
to approximate the spatial derivative of the $x$ component, $\textbf{F}^x$, of the physical flux appearing in \eqref{eq:conserv_PDE}.\footnote{In the following, we restrict the presentation to one dimension. A multidimensional scheme is obtained by considering fluxes in each direction separately and adding them to the r.h.s.} In the formula above $F^x$ is one of the components of $\textbf{F}^x$ with $F^x_i=F^x(x_i)$, $\hat f_{i\pm1/2}$ are the numerical fluxes at the cell interfaces and $h$ is the spatial grid spacing.

In order to construct a flux-limiter, the numerical fluxes in \eqref{eq:spatial_disc} have to be split into two contributions in the usual way \cite{Toro:1999}:
\begin{equation}
 \label{eq:num_flx_split}
  \hat f_{i\pm1/2} = \theta_{i\pm1/2} \hat f^{\,\mathrm{HO}}_{i\pm1/2} + 
  (1-\theta_{i\pm1/2}) \hat f^{\,\mathrm{LO}}_{i\pm1/2}.
\end{equation}
The quantities entering the r.h.s of \eqref{eq:num_flx_split} read: 
\begin{itemize}
 \item $f^{\,\mathrm{HO}}$ is a high-order (HO) unstable numerical flux used in regions where the numerical solution is smooth.
 \item $f^{\,\mathrm{LO}}$ is a high- or low-order (LO) stable numerical flux used in regions where the numerical solution develops non-smooth features. Notice that in order to be consistent with previous work \cite{Doulis:2022vkx,Doulis:2024aew}, we keep the notation LO although in the following we use exclusively a stable high-order scheme to approximate $f^{\,\mathrm{LO}}$. 
 \item $\theta \in [0,1]$ is a continuous weight function indicating how much from the numerical fluxes $f^{\,\mathrm{HO}}$ and $f^{\,\mathrm{LO}}$ to use at every instance.
\end{itemize}

Specifically, the flux $\hat f^{\,\mathrm{HO}}$ is constructed using the Rusanov Lax-Friedrichs flux-splitting technique, with reconstruction carried out on the characteristic fields  \cite{Mignone:2010br,Bernuzzi:2016pie}. A fifth-order central unfiltered stencil is always used for reconstruction. The flux $\hat f^{\,\mathrm{LO}}$ is estimated using the Local Lax-Friedrichs central scheme, where reconstruction is performed directly on the primitive variables  \cite{Thierfelder:2011yi}. Primitive reconstruction is always performed with the fifth-order weighted-essentially-non-oscillatory finite difference scheme WENOZ \cite{Borges:2008a}. With this choice, originally introduced in \cite{Doulis:2022vkx}, we generalise the traditional notion of a flux-limiter where $\hat f^{\,\mathrm{LO}}$ is always a LO monotone flux \cite{Toro:1999,Hesthaven:2017}. 
    
In order to compute the weight function $\theta$ we use the so-called \textit{entropy production function} $\nu$, a quantity that depends on the local entropy production. The relation between $\theta$ and $\nu$ reads
\begin{equation}
 \label{eq:def_theta}
  \theta_{i\pm1/2} = 1- \frac{1}{2} (\nu_i + \nu_{i\pm1}).
\end{equation}   

The entropy production function $\nu$ is computed from the specific entropy $s$ (entropy per unit mass) of the system under investigation using the second law of thermodynamics~\cite{Guercilena:2016fdl}: 
\begin{equation}
  \mathcal{R} = \nabla_\mu (s\,\rho\,u^\mu) \geq 0,
\end{equation}
where $\mathcal{R}$ is the so-called entropy residual and provides an estimation of the amount of entropy produced by the system under study. Following \cite{Guercilena:2016fdl}, the above expression can be written in terms of the time and spatial derivatives of the specific entropy as
\begin{equation}
 \label{eq:phys_entrp_resid_GRR}
 \mathcal{R} = \frac{\rho W}{\alpha}
 \left(\partial_t s + \mathrm{v}^i \partial_i s \right).
\end{equation}
Following the argumentation in \cite{Doulis:2022vkx}, we suppress the multiplication factor $\frac{\rho W}{\alpha}$ and replace $\mathcal{R}$ by
\begin{equation}
  \label{eq:phys_entrp_resid}
  R = \partial_t s + \mathrm{v}^i \partial_i s.
\end{equation}

Finally, we use the rescaled entropy residual $R$ defined above to define the \textit{entropy production function} as
\begin{equation}
 \label{eq:nu_E}
  \nu_E = c_E |R|,
\end{equation}
where $c_E$ is a tunable constant used to scale the absolute value of $R$. In the present work there is no need to adjust $c_E$; its value is always set to unity $c_E=1$. 

From its definition the parameter $\theta$ cannot be larger than the unity. Therefore, in order to ensure that the rhs of \eqref{eq:def_theta} does not exceed $[0,1]$ we have to impose the maximum value of $\nu_{\rm max}=1$ on $\nu$. Thus, the entropy production function entering \eqref{eq:def_theta} is given by 
\begin{equation}
 \label{eq:nu}
  \nu = \min\left[\nu_E, 1\right].
\end{equation}

The system of equations \eqref{eq:conserv_PDE} is solved numerically using the finite differencing code BAM \cite{Brugmann:2008zz,Thierfelder:2011yi,Dietrich:2015iva,Bernuzzi:2016pie} of which the EFL method is a module. For more information about the way that the EFL method is implemented in BAM the interested reader is referred to \cite{Doulis:2022vkx}. The value of the Courant-Friedrichs-Lewy condition has been set to 0.25 for all runs.

\section{Initial data}
\label{sec:ID}

In the present section, we show how to involve the EFL method in the eccentricity reduction of BNS initial data with SGRID. In order to demonstrate the wide applicability of the EFL method in the eccentricity reduction of such data, we consider in the following two different methods of reducing the eccentricity with the SGRID code. The first \cite{Dietrich:2015pxa} is based on an iterative algorithm that monitors and varies the initial radial velocity $\upsilon_r$ and eccentricity $e$ of the binary, while the second \cite{Tichy:2019ouu} uses also the initial radial velocity $\upsilon_r$ but replaces the eccentricity with the orbital angular velocity $\Omega$. The motivation for this lies in the latter method's enhanced ability to generate initial data for more extreme BNS configurations---such as those involving NSs of higher mass or/and spin---where the traditional approach may face limitations. Both approaches show similar behaviour when used in less extreme BNS scenarios \cite{Tichy:2019ouu}. 

\subsection{$\upsilon_r$ and $e$ driven eccentricity reduction}
\label{sec:BNSdata}

We start with the ``original'' SGRID eccentricity reduction algorithm \cite{Dietrich:2015pxa}. The initial data is constructed with the pseudospectral SGRID code \cite{Dietrich:2015pxa,Tichy:2006qn,Tichy:2009yr,Tichy:2009zr}, which makes use of the constant rotational velocity approach \cite{Tichy:2012rp,Tichy:2011gw} to construct spinning BNS configurations in hydrodynamical equilibrium. In accordance with \cite{Dietrich:2015pxa,Kyutoku:2014yba}, eccentricity reduced initial data are constructed through an iterative procedure of monitoring and varying the binary's initial radial velocity and eccentricity \cite{Dietrich:2015pxa}. As an initial guess, a quasi-equilibrium configuration in the usual quasi-circular orbit is employed with residual eccentricity of the order of $e \sim 10^{-2}$. The iterative procedure consist of the following steps: 
\begin{itemize}
\item[i)] Compute the initial data from the initial guess for the radial velocity and eccentricity.
\item[ii)] Evolve the data with BAM for 2-3 orbits.\footnote{We perform the dynamical evolution using $n=96$ points, see \autoref{tab:BNS_grid}.}
\item[iii)] Measure the eccentricity $e$ from the proper distance as described in \cite{Dietrich:2015pxa}.
\item[iv)] Re-compute the initial data with adjusted radial velocity and eccentricity.
\item[v)] Repeat steps ii) to iv) until the eccentricity cannot be further reduced.
\end{itemize}
In the following, each cycle of the type i)-iii) will be called an \textit{iteration}. Usually, it takes 3-4 iterations to reach the lowest eccentricity. Notice that the EFL method is involved in the ii) step of the above algorithm.

The above algorithm has been designed to fit the proper distance $d$ between the NSs centers, resulting from step ii), to the ansatz:
\begin{equation}
 \label{eq:d_fit}
  d(t) = S_0 + A_0 t + \frac{1}{2} A_1 t^2 - \frac{B}{\omega_f} \cos(\omega_f t + \phi),  
\end{equation} 
where $S_0, A_0, A_1, B, \omega_f$ and $\phi$ are fit parameters. By comparing \eqref{eq:d_fit} to the expected Keplerian orbits, under the assumption that the eccentricity $e$ is small for quasi-circular orbits, we compute the remaining eccentricity in the binary:
\begin{equation}
 \label{eq:eccentricity}
  e = \frac{B}{\omega_f d_0}
\end{equation}
and the necessary corrections to the radial velocity and eccentricity required to reduce the eccentricity in the next iteration:
\begin{equation}
 \label{eq:corr_BNSdata}
  \delta \upsilon_r = -B \sin\phi, \quad \delta e = \frac{B \cos\phi}{\omega_f d_0} (1 - e),
\end{equation}
where $d_0 \equiv d(0)$. 

To exemplify the use of the above procedure we will construct eccentricity reduced initial data for the non-spinning equal mass BNS configuration \BAM{95} \cite{Dietrich:2017aum} using the EFL method. This initial data is characterised by the parameters of \autoref{tab:bam95_ID}, namely by the Arnowitt-Deser-Misner (ADM) mass-energy $M_{\rm ADM}$, the angular momentum $J_0$, the baryonic mass $M_b$ and the dimensionless GW circular frequency $M \Omega_0$.

\begin{table}[t]
 \centering    
 \caption{\BAM{95} initial data. Columns: name; EoS; number of orbits; binary mass; rest-mass; ADM mass; angular momentum; GW frequency.\\}
   \begin{tabular}{ccccccccc}        
    \hline
    \hline
    Name & EoS & orbits & $M$ & $M_b$ & $M_\mathrm{ADM}$ & $J_0$ & $2M\Omega_0$\\
    \hline
    \BAM{95} & SLy & 10 & 2.700 & 2.989 & 2.678 & 7.686 & 0.038 \\
    \hline
    \hline
   \end{tabular}
 \label{tab:bam95_ID}
\end{table} 

Applying the aforedescribed reduction algorithm to \BAM{95}, we managed to reduce the eccentricity successively in each iteration down to the value $\sim 7\times10^{-4}$. The actual numerical values of the quantities involved in the reduction are shown in \autoref{tab:bam95_ecc_red}.

\begin{table}[t]
 \centering    
 \caption{\BAM{95} eccentricity reduction procedure. Columns: remaining eccentricity $e$ in the binary as measured at step iii); initial eccentricity $e_0$ and radial velocity $\upsilon_r$ used as input to SGRID at step i) or iv); corrections to the eccentricity and radial velocity as measured at step iii).\\}
   \begin{tabular}{cccccc}        
    \hline
    \hline
    Iter & $e [10^{-3}]$  & $e_0 [10^{-3}]$  & $\upsilon_r [10^{-3}]$  & $\delta e [10^{-3}]$  & $\delta \upsilon_r [10^{-3}]$  \\
    \hline
    1    & 10.14   & 0       & 0         & -4.70        & -1.54  \\
    2    & 3.12    & -4.70   & -1.54     & -1.66        & 0.38   \\
    3    & -0.76   & -6.36   & -1.16     & 0.21         & 0.15   \\
    \hline
    \hline
   \end{tabular}
 \label{tab:bam95_ecc_red}
\end{table}

\begin{figure}[h]
  \includegraphics[width=0.48\textwidth]{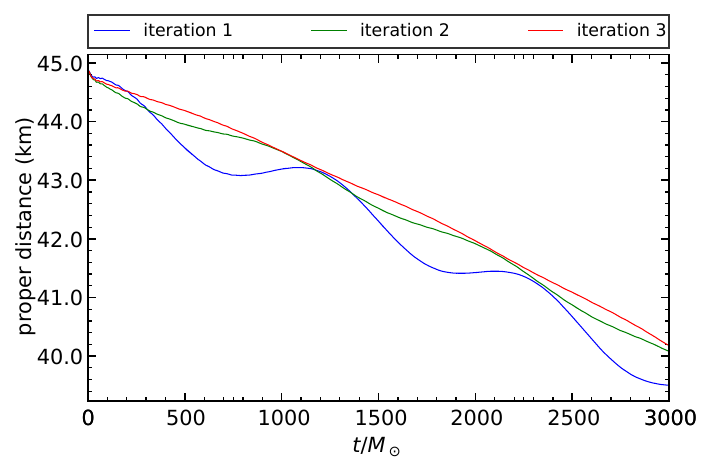}
  \caption{\BAM{95} proper distance as a function of time. As we proceed through the reduction procedure, the oscillations between the stars reduce and consequently the eccentricity of the initial data.}
  \label{fig:bam95_sgrid_iter_d_proper}
\end{figure}

A visualisation of the above eccentricity reduction can be found in \autoref{fig:bam95_sgrid_iter_d_proper}. Therein, on each hypersurface, the proper distance $d$ between the centers of the two NSs is plotted as a function of time for the first $\sim$3 orbits of the inspiral. The reduction of the oscillations in the separation between the stars is apparent as we proceed from the initial to the final iteration, indicating a decline of the eccentricity in the initial data. 

\subsection{$\upsilon_r$ and $\Omega$ driven eccentricity reduction}
\label{sec:DNSdata}

In order to enable the study of more extreme mass and spin BNS configurations, a modified version of  the algorithm presented in \autoref{sec:BNSdata} was introduced in \cite{Tichy:2019ouu}, where the eccentricity $e$ is replaced by the orbital angular velocity $\Omega$ of the binary. As an initial guess for the radial and orbital angular velocity we set $\upsilon_r=0$ and consider a post-Newtonian estimate for $\Omega$, respectively. The resulting iterative procedure resembles the one described in \autoref{sec:BNSdata}: 
\begin{itemize}
\item[i)] Compute the initial data from the initial guess for $\upsilon_r$ and $\Omega$.
\item[ii)] Evolve the data using BAM with the EFL method for 2-3 orbits.\footnote{Also here the resolution used is $n=96$ points, see \autoref{tab:BNS_grid}.}
\item[iii)] Measure the eccentricity $e$ from the proper distance as described in \cite{Tichy:2019ouu}.
\item[iv)] Re-compute the initial data with adjusted radial and orbital angular velocity.
\item[v)] Repeat steps ii) to iv) until the eccentricity cannot be further reduced.
\end{itemize}
Again each cycle of the type i)-iii) is called an \textit{iteration}. With the above algorithm we want to estimate the parameters of the ansatz \eqref{eq:d_fit} from the numerically computed proper distance of step ii). The relation \eqref{eq:eccentricity} holds also here but \eqref{eq:corr_BNSdata} has to be replaced with
\begin{equation}
 \label{eq:corr_DNSdata}
  \delta \upsilon_r = -B \sin\phi, \quad \delta \Omega = \frac{B \omega_f \cos\phi}{2 \Omega d_0}.
\end{equation}

In the following, we will use the above procedure to construct eccentricity reduced initial data for the spinning unequal mass BNS configuration MPA1q1.6 with parameters given in \autoref{tab:MPA1_ID}. The masses and dimensionless spins of the individual NSs of the binary are $(M_A, M_B) = (1.868, 1.173)$ and $(\chi_A, \chi_B) = (0.189, 0.157)$, respectively. The mass ratio of the binary is $q=1.593$ indicating a moderate mass asymmetry. 

\begin{table}[t]
 \centering    
 \caption{MPA1q1.6 initial data. Columns: name; EoS; number of orbits; binary mass; rest-mass; ADM mass; angular momentum; GW frequency.\\}
   \begin{tabular}{ccccccccc}        
    \hline
    \hline
    Name & EoS & orbits & $M$ & $M_b$ & $M_\mathrm{ADM}$ & $J_0$ & $2M\Omega_0$\\
    \hline
    MPA1q1.6 & MPA1 & 6 & 3.041 & 3.430 & 3.017 & 9.380 & 0.052 \\
    \hline
    \hline
   \end{tabular}
 \label{tab:MPA1_ID}
\end{table} 

\autoref{tab:MPA1_ecc_red} contains the numerical values of the quantities involved in the reduction. The eccentricity gradually reduces after each iteration and after the final fourth iteration reaches its minimum value of $\sim 2 \times 10^{-3}$. Notice that the eccentricity reduction of the initial data for MPA1q1.6 require an additional iteration compared to \BAM{95}. Most probably this happens because the reduction procedure for MPA1q1.6 is more demanding because of the spin and mass asymmetry of the data.     

\begin{table}[t]
 \centering    
 \caption{MPA1q1.6 eccentricity reduction procedure. Columns: remaining eccentricity $e$ in the binary as measured at step iii); initial orbital angular velocity $\Omega_0$ and radial velocity $\upsilon_r$ used as input to SGRID at step i) or iv); corrections to the orbital angular velocity and radial velocity as measured at step iii).\\}
   \begin{tabular}{cccccc}        
    \hline
    \hline
    Iter & $e [10^{-2}]$  & $\Omega_0 [10^{-3}]$  & $\upsilon_r [10^{-3}]$  & $\delta \Omega [10^{-5}]$  & $\delta \upsilon_r [10^{-3}]$  \\
    \hline
    1    & -14.20  & 8.89   &  0      & -24.81    &  8.16  \\
    2    & -6.26   & 8.64   &  8.16   & -11.70    & -7.09  \\
    3    &  1.46   & 8.52   &  1.08   &  0.13     & -2.98  \\
    4    &  0.21   & 8.53   & -1.91   &  0.08     & -0.48  \\
    \hline
    \hline
   \end{tabular}
 \label{tab:MPA1_ecc_red}
\end{table}

\begin{figure}[h]
  \includegraphics[width=0.48\textwidth]{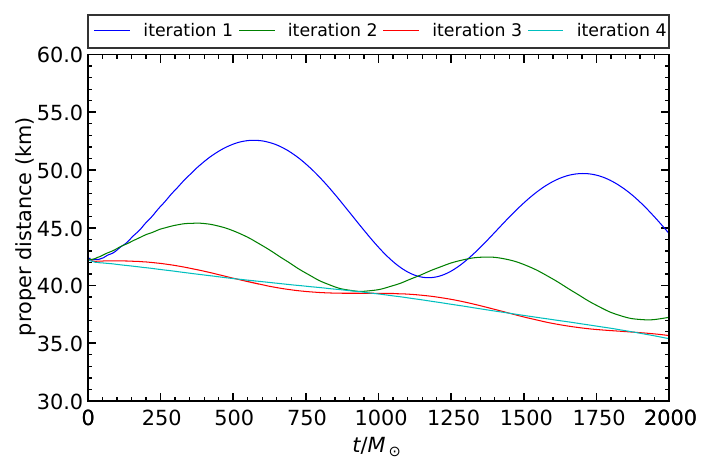}
  \caption{MPA1q1.6 proper distance as a function of time. As we proceed through the reduction procedure, the initial quite large oscillations between the stars gradually reduce and essentially disappear at the last iteration.}
  \label{fig:MPA1_sgrid_iter_d_proper}
\end{figure}

The qualitative behaviour of the eccentricity during the reduction process can be visualised by plotting the proper distance, resulting from step ii), as a function of time. This is depicted in \autoref{fig:MPA1_sgrid_iter_d_proper}, where the dynamical behaviour of the proper distance is plotted at each iteration for the first $\sim3$ orbits. Notice that the initial oscillations here are considerably larger than the ones of \BAM{95} depicted in \autoref{fig:bam95_sgrid_iter_d_proper}, most probably because of the spin and mass asymmetry of the data. Consequently, we need one more iteration to smooth out the oscillations.

\section{Binary neutron star evolutions}
\label{sec:BNS}

We present a couple of BNS simulations where the EFL method is used both in the eccentricity reduction of the initial data, see \autoref{sec:ID}, and the dynamical evolution. 

\subsection{\normalsize\BAM{95}}
\label{sec:bam95_BNS}

\subsubsection{Numerical setup}
\label{sec:bam95_setup}

\begin{figure*}[t]
  \centering 
  \begin{tabular}[c]{ccc}
  \includegraphics[width=0.32\textwidth]{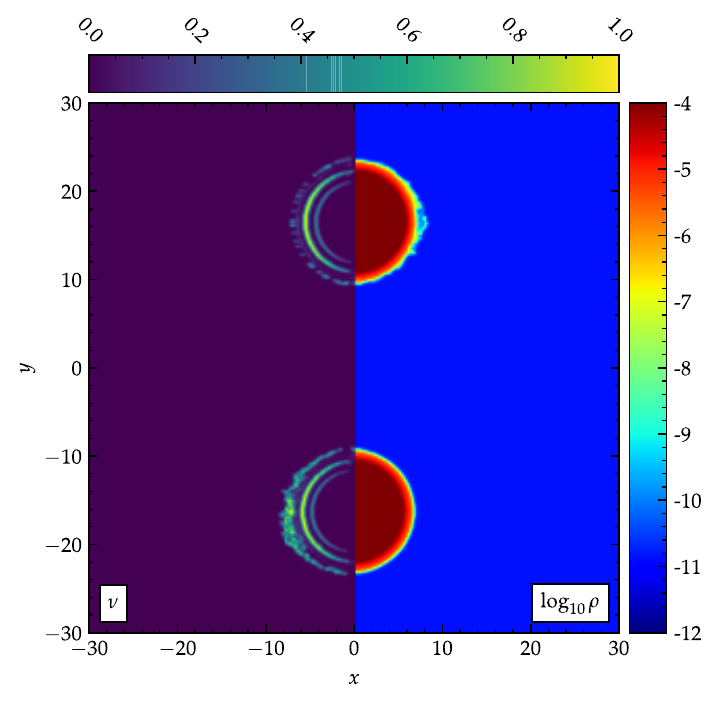}
  \includegraphics[width=0.32\textwidth]{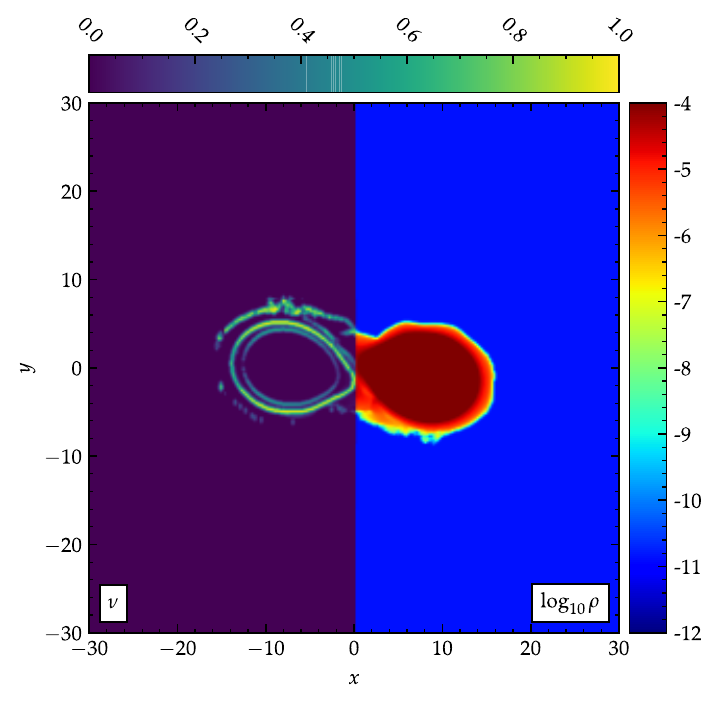}
  \includegraphics[width=0.32\textwidth]{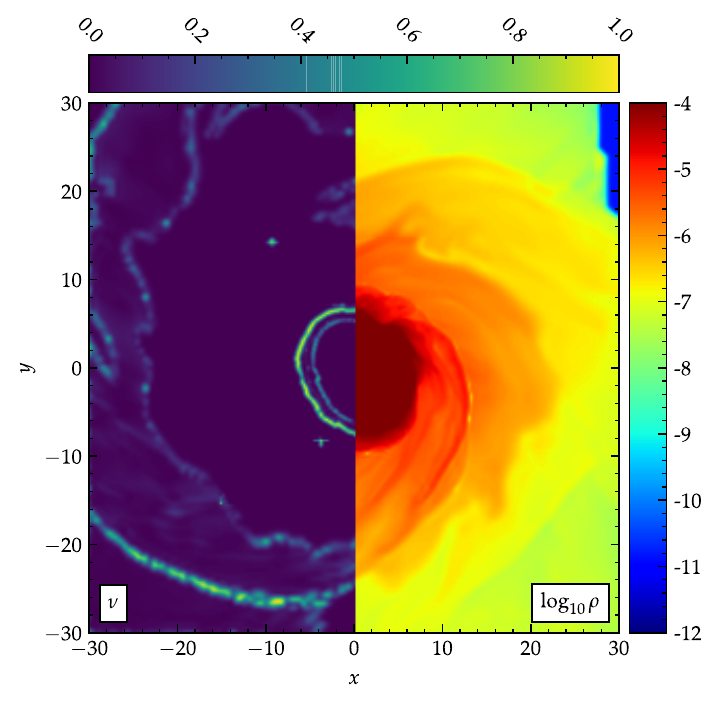}
  \end{tabular}
  \caption{Two-dimensional hybrid plots depicting the entropy production and rest-mass density profiles across different stages of the pure EFL \BAM{95} simulation. Left: inspiral. Middle: merger. Right: post-merger.}
  \label{fig:hybrid_plots_bam95}
\end{figure*}

In the following, we study the dynamics of the BNS initial data constructed in \autoref{sec:BNSdata} for \BAM{95} \cite{Dietrich:2017aum}. This configuration is similar to \BAM{97} \cite{Dietrich:2018phi}, which was studied within the EFL framework in \cite{Doulis:2022vkx}. The main difference between these two BNS configurations lies in the way their initial data have been constructed: \BAM{97} uses the Lorene library \cite{Gourgoulhon:2000nn} while \BAM{95} uses initial data constructed with the SGRID code \cite{Tichy:2012rp}.

We chose this BNS simulation because it enables us to study the effect of using the EFL method in the eccentricity reduction of the initial data. In all the BNS simulations presented in \cite{Doulis:2022vkx} the EFL method was used only in the dynamical evolution of Lorene initial data. With \BAM{95} the EFL method can be involved in the eccentricity reduction 
of the initial data through the eccentricity reduction process. 

In the present work, we use different numerical schemes to compute the numerical fluxes \eqref{eq:spatial_disc} involved in the construction of the initial data and the dynamical evolution. The main reason for doing this is to study if the effect of using the EFL method will lead to higher accuracy and better convergence properties for our numerical solutions. In \cite{Dietrich:2017aum} the HO-LLF method was used both in the eccentricity reduction of the initial data for \BAM{95} and in the subsequent evolution with BAM. Here, we use the EFL method to conduct two additional studies of the \BAM{95} configuration. In the first, the HO-LLF method is used in the initial data eccentricity reduction and the EFL method in the subsequent BAM evolution. In the second, both the eccentricity reduction of the initial data and the BAM evolution is conducted with the EFL method. The aforementioned different combinations of numerical schemes used to study \BAM{95} in the present work are summarised in \autoref{tab:BNS_ecc_bam95}. 

\begin{table}[h]
 \centering    
 \caption{\BAM{95} case studies. Columns: name of \BAM{95} case study; method used in the 
eccentricity reduction of the SGRID initial data; method used to evolve the initial data with BAM; reconstruction scheme used in the simulations; eccentricity.\\}
   \begin{tabular}{ccccc}        
    \hline
    \hline
    Name          & Ecc. Red.    & BAM     & Reconstruction  & $e [10^{-3}]$  \\
    \hline
    pure HO-LLF   & HO-LLF       & HO-LLF  & WENOZ           & 0.4  \\
    hybrid EFL    & HO-LLF       & EFL     & WENOZ           & 0.4  \\
    pure EFL      & EFL          & EFL     & WENOZ           & 0.7  \\
    \hline
    \hline
   \end{tabular}
 \label{tab:BNS_ecc_bam95}
\end{table}

The initial data for \BAM{95} is evolved in the three different combinations of \autoref{tab:BNS_ecc_bam95} with the WENOZ reconstruction scheme. WENOZ reconstruction is used because the results in \cite{Doulis:2022vkx} strongly indicate that WENOZ is the best performing reconstruction scheme from those available. For each case study three different grid resolutions are considered. The grid specifications for all \BAM{95} runs are the same and are reported in \autoref{tab:BNS_grid}. The atmosphere setting for all simulations are $f_{\rm atm}=10^{-11}$ and  $f_{\rm thr}=10^2$. The metric is evolved with the Z4c scheme. Standard radiative boundary conditions are used for all BNS simulations. The behaviour and the convergence properties of the conserved quantities of the present \BAM{95} runs are similar to those of \BAM{97} presented in \cite{Doulis:2022vkx}. Therefore, in the following, for the sake of presentational brevity and clarity, we solely focus on the qualitative behaviour of the entropy production and the gravitational wave analysis of our \BAM{95} runs. 

\begin{table}[h]
 \centering    
 \caption{Grid configurations for all the \BAM{95} case studies. Columns (left to right): refinement levels; minimum moving level index; number of points per direction in fixed levels; number of points per direction in moving levels; resolution per direction in the finest level $l=L-1$; resolution per direction in the coarsest level $l = 0$.\\}
   \begin{tabular}{cccccc}        
    \hline
    \hline
    $L$ & $l^\mathrm{mv}$ & $n^\mathrm{fix}$ & $n$ & $h_{L-1}$ & $h_0$ \\
    \hline 
     7 & 2 & 160 & 64  & 0.235  & 15.040  \\
     7 & 2 & 240 & 96  & 0.157  & 10.027   \\
     7 & 2 & 320 & 128 & 0.118  & 7.520   \\
    \hline
    \hline
   \end{tabular}
 \label{tab:BNS_grid}
\end{table} 

\subsubsection{Qualitative behaviour of the entropy production}
\label{sec:bam95_hybrid_plots}

The entropy production function $\nu$ plays central role in our method. Hence, it is of great interest to study its behaviour during the evolution of BNS merger simulations. In the following, we discuss the two-dimensional entropy production profiles of  the ten-orbit simulation \BAM{95}. 

In \autoref{fig:hybrid_plots_bam95} we present two-dimensional hybrid plots depicting the entropy production function $\nu$ and rest-mass density $\rho$ profiles at different stages of the pure EFL simulation. (The hybrid EFL case shows similar behaviour.) The different panels show (from left to right) selected snapshots of the inspiral, merger and post-merger stages, respectively. Taking a closer look at the $\nu$ profile, we observe that during the inspiral and merger the EFL method is, as expected \cite{Doulis:2022vkx,Doulis:2024aew}, only activated at the surface of each NS. The same can be also observed in the post-merger regime, however, here also regions in the exterior of the remnant NS are flagged by the EFL method. Judging from the corresponding rest-mass density plots, the observed flagging in the exterior is actually capturing the outward dynamics of the spiral density waves emitted by the rotating remnant of the two inspiraling NSs. This fact shows the effectiveness of the EFL method in locating any non-smooth features developing in the numerical domain. 

In summary, $\nu$ appears to be concentrated quite uniformly along the NS surfaces during inspiral and merger. The post-merger stage shows spiral arms with moderate complexity. Because of its equal-mass, non-spinning nature, \BAM{95} maintains symmetry in both $\nu$ and $\rho$ profiles across all phases. The post-merger density waves exhibit well-defined spiral patterns that are minimally disrupted. 

\subsubsection{Gravitational wave analysis}
\label{sec:gw_analysis_bam95}

\begin{figure*}[t]
  \begin{tabular}[c]{ccc}
  \vspace{-3mm}
  \includegraphics[width=0.6\textwidth]{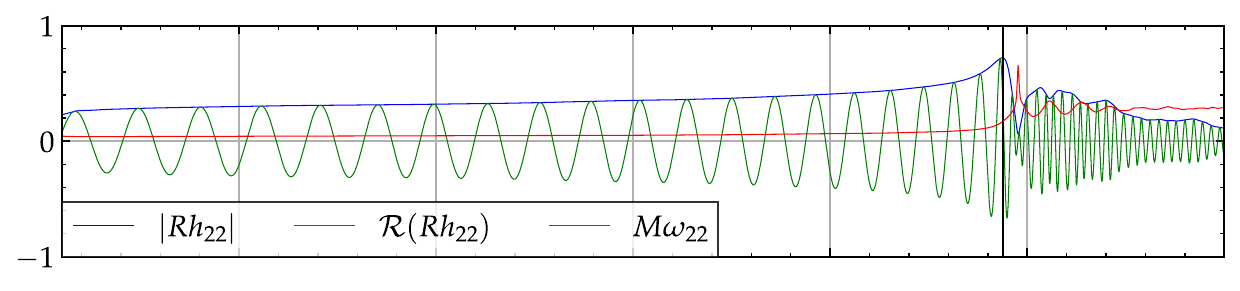}\\
  \vspace{-3mm}
  \includegraphics[width=0.6\textwidth]{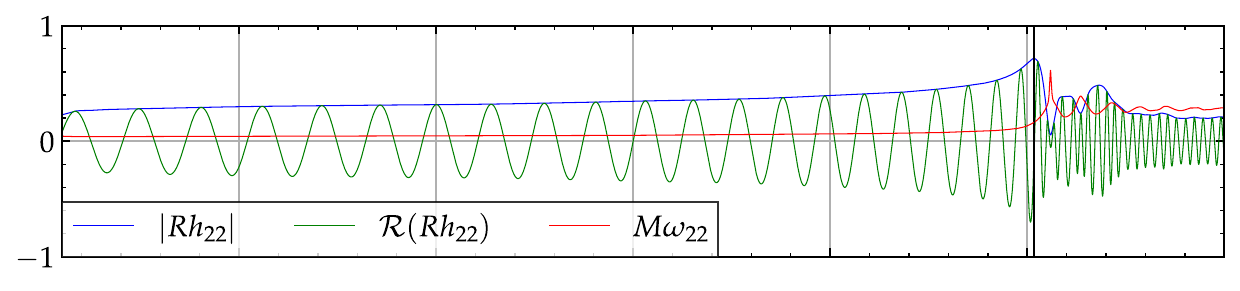}\\
  \quad\,\includegraphics[width=0.615\textwidth]{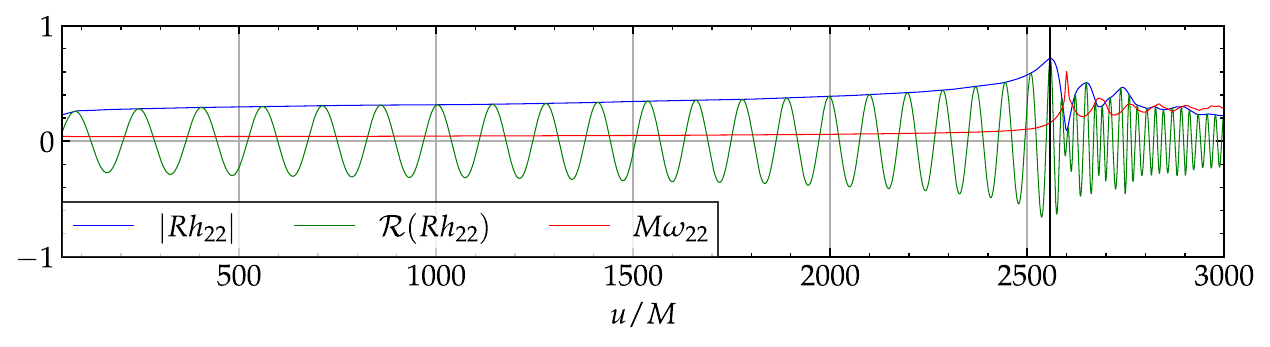}
  \end{tabular}
  \put(-290,95){pure HO-LLF}
  \put(-290,28){hybrid EFL}
  \put(-290,-38){pure EFL}
  \caption{\BAM{95} waveforms. The amplitude (blue line), the real part (green line) and the instantaneous frequency $M\omega_{22}$ (red line) of the GW signal obtained from the ten-orbit \BAM{95} simulation using WENOZ reconstruction for the three case studies of \autoref{tab:BNS_ecc_bam95}: pure HO-LLF (top), hybrid EFL (middle), and pure EFL (bottom). The time of merger is defined as the first peak of the amplitude $A_{22}$ and is indicated by a black vertical line.}
  \label{fig:gw_bam95}
\end{figure*}

\begin{figure*}[t]
  \begin{tabular}[c]{ccc}
  \includegraphics[width=0.33\textwidth]{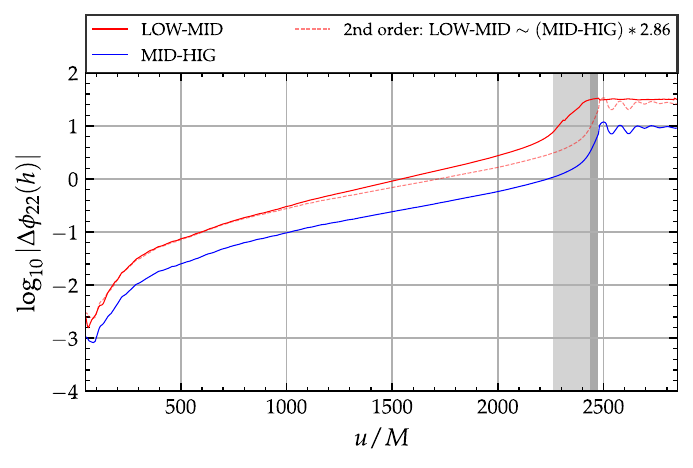}
  \includegraphics[width=0.33\textwidth]{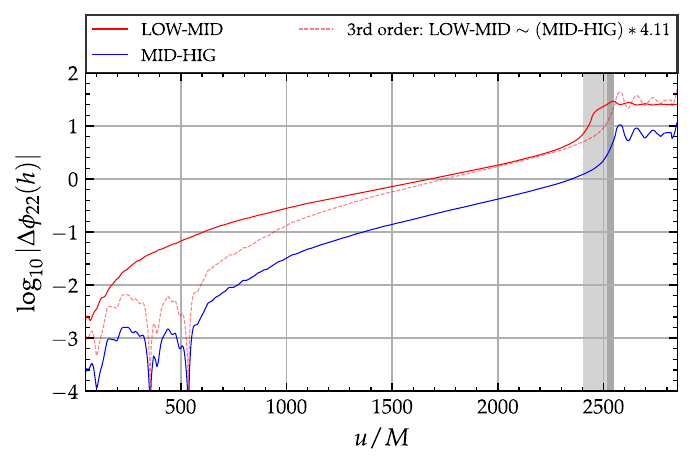}
  \includegraphics[width=0.33\textwidth]{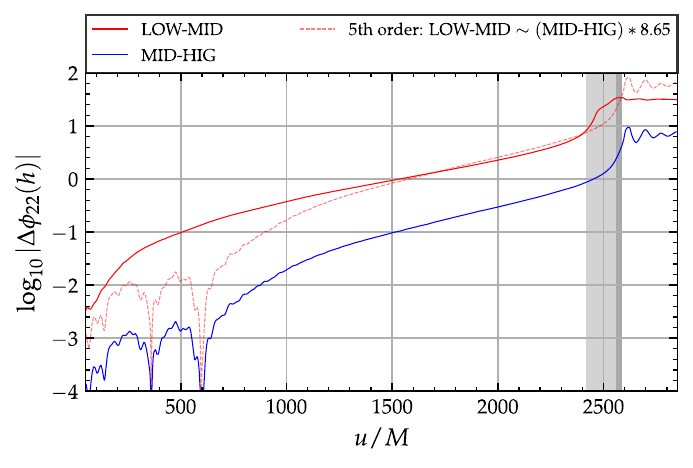}
  \end{tabular}
  \put(-485,31){\scriptsize pure HO-LLF}
  \put(-313,31){\scriptsize hybrid EFL}
  \put(-141,31){\scriptsize pure EFL}
  \caption{GW phase difference convergence rate study for the ten-orbit \BAM{95} simulation. Left: pure HO-LLF case. Middle: hybrid EFL case. Right: pure EFL case.}
  \label{fig:gw_phase_errors_bam95}
\end{figure*}

We study now the effect of using the \EFL{} method on the gravitational waveforms (GWs). Following \cite{Bernuzzi:2016pie,Doulis:2022vkx}, we use the curvature scalar $\Psi_4$ to compute the GWs on spheres at distance $r$ from the origin. First, spin weighted spherical harmonics are used to expand $\Psi_4$ into its modes $\psi_{\lm}$ and then the multipolar modes $h_{\lm}$ are reconstructed by solving the expression $\ddot{h}_{\lm} = \psi_{\lm}$ in the frequency domain \cite{Reisswig:2010di}. In terms of the amplitude $A_{\lm}$ and phase $\phi_{\lm}$ the different modes of $h_{\lm}$ are given by the expression
\begin{align}
  R\, h_{\lm} = A_{\lm} e^{- i \phi_{\lm}}.
  \label{eq:polar_representation_gw_strain}
\end{align}
In the present section, we plot our results against the retarded time 
\begin{align}
  u = t - R(r) - 2 M \log\left(\frac{R(r)}{2M}-1\right),
\end{align}
where $M$ is the total gravitational mass of the BNS system, $R(r) = r (1 + M/2r)^2$ is the radius in Schwarzschild coordinates and $r$ is the extraction radius. The time of merger $u_{\rm mrg}$ is estimated from the first peak of $A_{22}$ of the dominant $(\ell,m)=(2,2)$ mode 

The waveforms obtained from the ten-orbit inspiral \BAM{95} simulation for the three case studies of \autoref{tab:BNS_ecc_bam95} are presented in \autoref{fig:gw_bam95}. The wave train of all three waveforms after a first peak, indicating the time of merger, shows a more complicated structure that involves multiple denser distributed peaks and a slow amplitude decay. The instantaneous GW frequency $\omega_{22} = - \mathcal{I}(\dot h_{22} / h_{22})$ is also plotted (red line), which undergoes a sharp frequency increase near merger (a feature characteristic of a chirp-like signal). Notice that from top to bottom the time of merger is increasing, with the difference between pure HO-LLF and pure EFL being around $u_{\rm mrg} \approx 150 M$. Clearly, the use of the EFL method elongates the inspiral and increases the merger time. 

Next, self-convergence studies are performed based on simulations using $(n_i) = (64, 96, 128) \equiv (\textrm{LOW, MID, HIG})$ points per direction on the highest AMR level. The phase difference between pairs of resolutions is defined by the expression  
\begin{align}
  \Delta \phi^{(n_i,n_j)}_{\lm} = \phi^{(n_i)}_{\lm} - \phi^{(n_j)}_{\lm}\,.
  \label{eq:phase_diff}
\end{align} 
The experimental convergence rate $p$ is determined through the rescaling of the resulting differences by a factor $s$ that expresses the theoretically expected rate of decrease of these differences with increasing resolutions, provided of course that our simulations are converging. The definition of the factor $s$ follows \cite{Baumgarte:2010}: 
\begin{align}
   s(p, n_i, n_j, n_k) &= \frac{1 - (n_i/n_j)^p}{(n_i/n_j)^p - (n_i/n_k)^p},
     \label{eq:convergence_scaling}
\end{align}
where $n_i < n_j < n_k$. 

\autoref{fig:gw_phase_errors_bam95} shows a self-convergence study of the waveform phase differences for the \BAM{95} simulations. The three panels (from left to right) correspond to results obtained with the pure HO-LLF, hybrid EFL, and pure EFL schemes, respectively. The pure HO-LLF result \cite{Dietrich:2017aum} (left panel) serves as a reference for comparison with the case studies involving the \EFL{} method. Solid lines in \autoref{fig:gw_phase_errors_bam95} represent phase differences between runs with consecutively increasing resolutions; dashed lines correspond to rescaled differences of MID-HIG differences, where the scaling factor is computed from \eqref{eq:convergence_scaling} using an integer convergence rate $p$; the gray shaded regions mark the differences in merger times between runs with increasing resolution.

As expected the phase differences between runs with increasing resolutions decrease for all simulations. A feature that indicates that all schemes are capable of providing results that converge to some order. This observation is further supported by the fact that the difference between merger times is reducing, indicated by the narrowing of the gray shaded regions, with increasing resolution. Notice that, in accordance with \autoref{fig:gw_bam95}, the use of the EFL method clearly increases the time of merger. 

All convergence series demonstrate a clear convergence behaviour: the pure HO-LLF second-order, the hybrid EFL third-order, and the pure EFL fifth-order convergence. The expected rate of convergence is 5th-order for all cases. We observe that optimal convergence is achieved only in the pure EFL case. Comparing the results obtained with the \EFL{} method to the ones obtained with the HO-LLF method, we find that the use of the EFL method leads to higher convergence rates and to generically smaller absolute differences MID-HIG at merger. The pure HO-LLF case (left panel) yields a clear second-order convergence, consistent with the results in \cite{Dietrich:2017aum}, and phase difference MID-HIG at merger a factor $\sim 2$ larger than the pure EFL case (right panel). We also note that in the early post-merger phase the convergence rate for the EFL cases is higher (third-order) and the difference MID-HIG smaller. An observation that unveils the robustness of the EFL method and its ability to capture satisfactorily the violent dynamics of the remnant NS. 

\begin{figure}[h]
  \centering
  \begin{tabular}[c]{c}
  \includegraphics[width=0.45\textwidth]{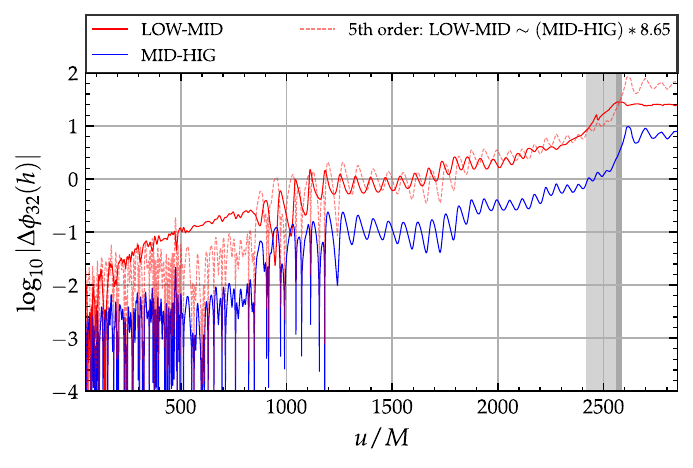}\\
  \includegraphics[width=0.45\textwidth]{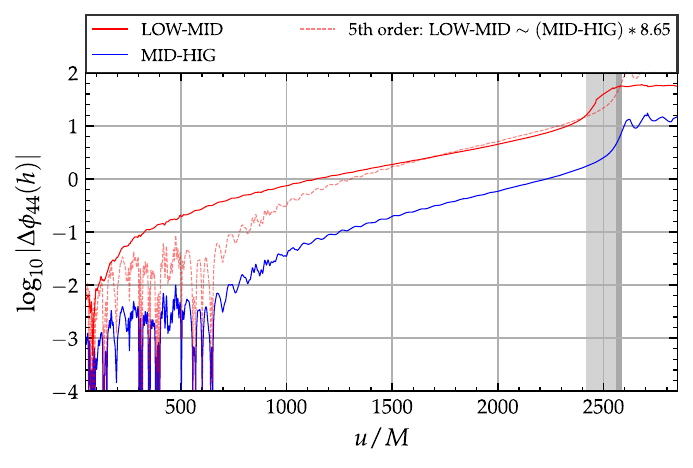}
  \end{tabular}
  \put(-193,120){pure EFL}
  \put(-193,-38){pure EFL}
  \caption{GW phase difference convergence rate study for higher modes of the pure EFL ten-orbit \BAM{95} simulation with WENOZ reconstruction. Top: $(3,2)$-mode. Bottom: $(4,4)$-mode.
  \label{fig:gw_phase_errors_bam95_higher_modes}}
\end{figure}

Given this clear higher-order convergence pattern for the dominant $(2,2)$-mode of the pure EFL case, it is natural to also investigate the convergence of other (subdominant) modes of $h_{\lm}$ with $\ell>2$. \autoref{fig:gw_phase_errors_bam95_higher_modes} shows a convergence study of the $(\l,m) = (3,2)$ and $(4,4)$ modes. Also in this case, the phase differences show a consistent decrease with increasing resolution and a clear fifth-order convergence behaviour. Furthermore, a third-order convergence pattern clearly holds through merger and in the early post-merger regime.

It is worth emphasizing that the \BAM{97} simulation, which utilized initial data generated with the Lorene library, achieved fourth-order convergence \cite{Doulis:2022vkx}. A key limitation of Lorene-based initial data is its inability to undergo eccentricity reduction, which can result in a higher level of residual orbital eccentricity in the initial conditions. In contrast, the \BAM{95} simulation employed initial data constructed with the SGRID code, which allows for eccentricity reduction. This critical capability, combined with the use of the EFL method during the eccentricity reduction process, likely enabled \BAM{95} to achieve a notably higher fifth-order convergence rate.

\begin{figure*}[t]
  \centering 
  \begin{tabular}[c]{ccc}
  \includegraphics[width=0.32\textwidth]{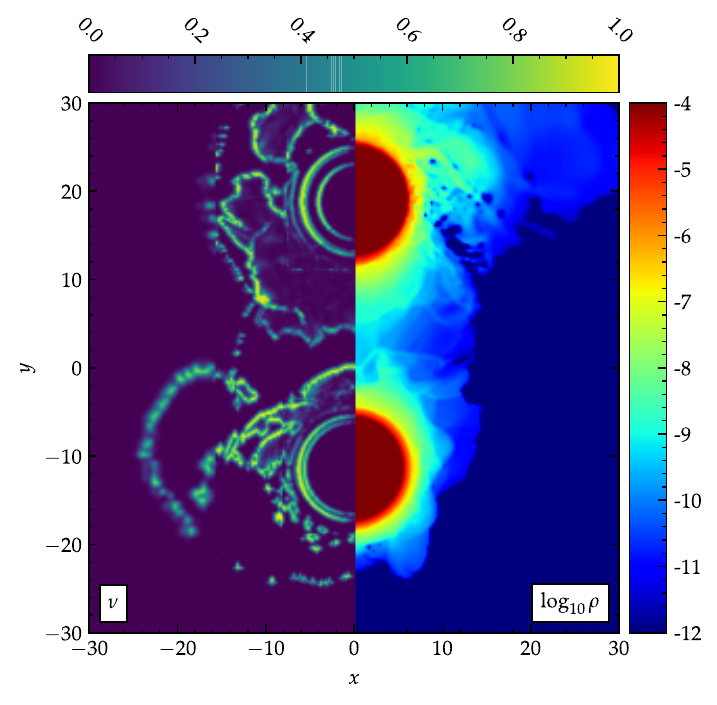}
  \includegraphics[width=0.32\textwidth]{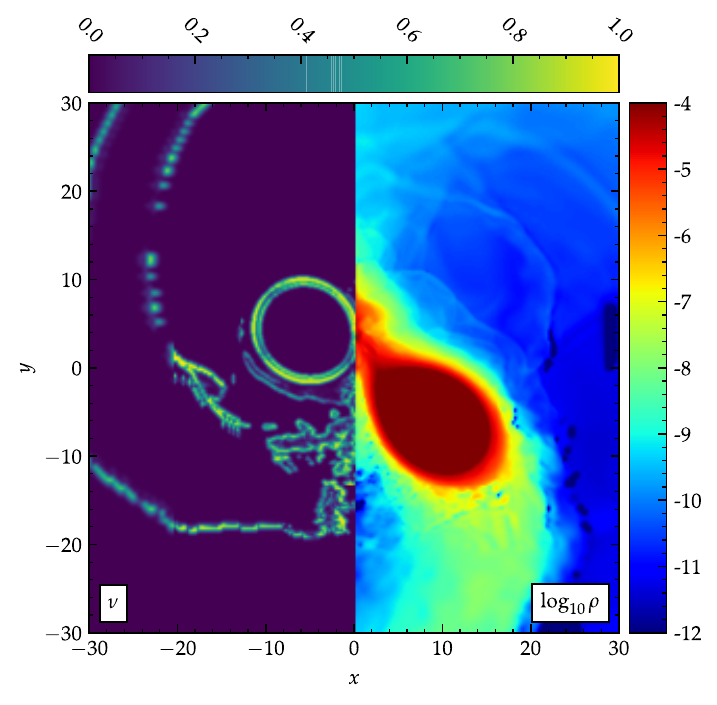}
  \includegraphics[width=0.32\textwidth]{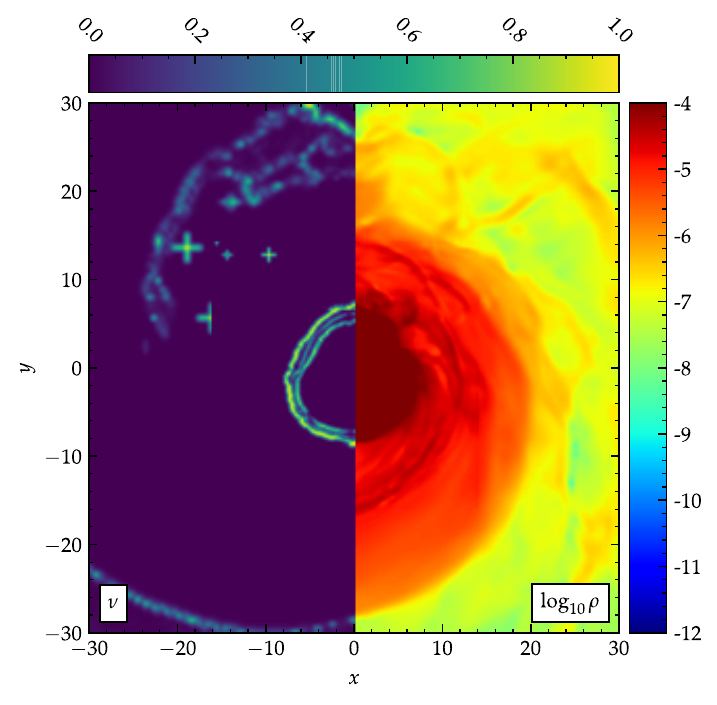}
  \end{tabular}
  \caption{Two-dimensional hybrid plots depicting the entropy production and rest-mass density profiles across different stages of the MPA1q1.6 simulation. Left: inspiral. Middle: merger. Right: post-merger.}
  \label{fig:hybrid_plots_MPA1}
\end{figure*}

\begin{figure*}[t]
  \vspace{-3mm}
  \includegraphics[width=0.6\textwidth]{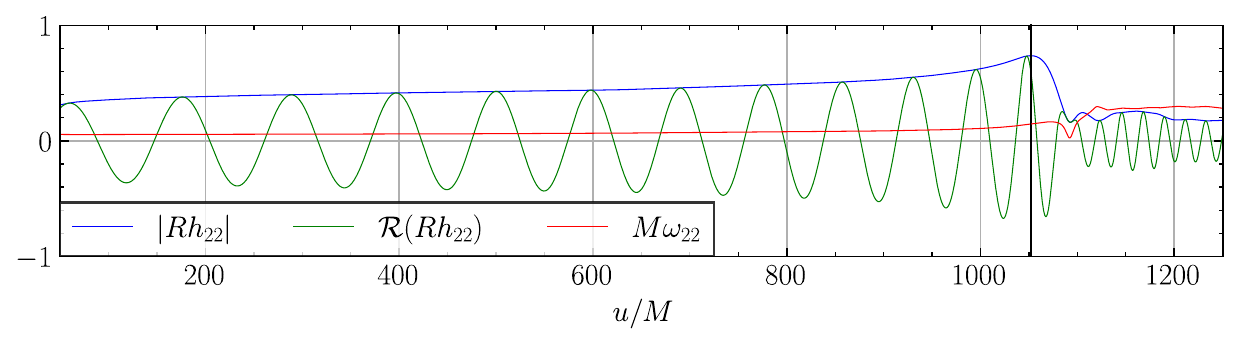}\\
  \caption{MPA1q1.6 waveform. The amplitude (blue line), the real part (green line) and the instantaneous frequency $M\omega_{22}$ (red line) of the GW signal obtained from the MPA1q1.6 simulation using WENOZ reconstruction for the case study of \autoref{tab:BNS_ecc_MPA1}. The time of merger is defined as the first peak of the amplitude $A_{22}$ and is indicated by a black vertical line.}
  \label{fig:gw_MPA1}
\end{figure*}

\subsection{MPA1q1.6}

\subsubsection{Numerical setup}
\label{sec:MPA1_setup}

The detailed case study presented in \autoref{sec:bam95_BNS} provides compelling evidence that the use of the pure EFL case results in GWs with significantly improved accuracy and enhanced convergence properties. Based on these findings, we focus exclusively on the pure EFL case in the subsequent analysis of the MPA1q1.6 configuration, as detailed in \autoref{tab:BNS_ecc_MPA1}. 

The previously discussed \BAM{95} configuration, along with all the BNS simulations reported in \cite{Doulis:2022vkx}, correspond to non-spinning, equal-mass setups. These studies have explicitly demonstrated that adopting the EFL method enhances both the accuracy and convergence properties of the simulations. The MPA1q1.6 configuration, however, is an unequal-mass binary neutron star system with spin, offering a valuable opportunity to evaluate the performance of the EFL method in a more complex scenario.

The initial data for the MPA1q1.6 configuration was eccentricity reduced using the EFL method as described in \autoref{sec:DNSdata}. The numerical setup for the dynamical evolution of this data is the same with the one used for \BAM{95} in \autoref{sec:bam95_BNS}. Details regarding the resolutions and the grid parameters used to define the numerical setup for the simulations discussed in the present section are provided in \autoref{tab:BNS_grid}.

\begin{table}[h]
 \centering    
 \caption{MPA1q1.6 case study. Columns: name; method used in the eccentricity reduction of the SGRID initial data; method used to evolve the initial data with BAM; reconstruction scheme used in the simulations; eccentricity.\\}
   \begin{tabular}{ccccc}        
    \hline
    \hline
    Name          & Ecc. Red.    & BAM     & Reconstruction  & $e [10^{-2}]$  \\
    \hline
    pure EFL      & EFL          & EFL     & WENOZ           & 0.2  \\
    \hline
    \hline
   \end{tabular}
 \label{tab:BNS_ecc_MPA1}
\end{table}

\subsubsection{Qualitative behaviour of the entropy production}
\label{sec:MPA1_hybrid_plots}

We start with an analysis of the two-dimensional entropy production profiles obtained from the MPA1q1.6 simulation. This will enable us to study the behaviour of the entropy production function, $\nu$, during different phases of the evolution which will provide us with crucial insights into the EFL method's effectiveness and the dynamics of the system. 

\autoref{fig:hybrid_plots_MPA1} illustrates a series of two-dimensional hybrid plots showcasing the entropy production function, $\nu$, alongside the rest-mass density, $\rho$, at various stages of the simulation. The panels in the figure present snapshots corresponding to three distinct phases of the simulation: the inspiral (left panel), merger (middle panel), and post-merger (right panel). A detailed examination of the $\nu$ profiles reveals that the EFL method is primarily activated near the surface of each NS during all phases of the evolution. This behaviour highlights the role of the method in identifying and addressing non-smooth features arising at the star's boundaries, where steep gradients in the hydrodynamical variables naturally occur. Moreover, we observe additional regions of activation in the exterior of the stellar surface. Upon comparing the entropy production profiles to the corresponding rest-mass density plots, it becomes evident that these features appear in the former because of the matter cloud surrounding the NSs. In addition, in the post-merger phase, this behaviour is enhanced by the outward-propagating spiral density waves emitted by the rapidly rotating remnant. These density waves are a direct consequence of the violent merger process, carrying angular momentum and energy outwards. 

By comparing with \autoref{fig:hybrid_plots_bam95}, it becomes apparent that in the MPA1q1.6 simulation, $\nu$ exhibits a more intricate pattern during inspiral, with additional structures potentially arising from unequal mass or spin effects. The post-merger stage reveals a more chaotic $\nu$ distribution in the remnant, reflecting stronger or more widespread outward dynamics. Being an unequal mass system with spin, MPA1q1.6 introduces asymmetry. This is evident in the non-uniform distribution of $\nu$ and the skewed shapes of density waves in the merger and post-merger phases. 

\subsubsection{Gravitational wave analysis}
\label{sec:gw_analysis_MPA1}

We analyse now the GWs resulting from the dynamical evolution of MPA1q1.6. The gravitational waveform derived from the MPA1q1.6 simulation is shown in \autoref{fig:gw_MPA1}. The waveform initially exhibits a periodic and smooth oscillatory pattern, characteristic of the inspiral phase. As the system evolves, the amplitude steadily increases due to the gradual tightening of the binary orbit. The merger event is marked by the first significant peak in the waveform, after which the wave train develops a more intricate structure. This post-merger stage displays multiple closely spaced peaks, which are indicative of the complex dynamics of the remnant. Notably, the waveform amplitude experiences a slow decay in this regime.

Additionally, the instantaneous GW frequency is plotted in red. This frequency profile undergoes a sharp increase near the time of merger, representing the characteristic chirp-like signal associated with binary systems nearing coalescence. The steep rise in $\omega_{22}$ reflects the rapid acceleration and strong interaction between the NSs as they approach their final collision. This analysis demonstrates the ability of the simulation to capture the transition from inspiral to post-merger dynamics with high fidelity, showcasing both the amplitude and frequency evolution of the emitted GWs.

\begin{figure}[h]
  \centering
  \begin{tabular}[c]{ccc}
    \includegraphics[width=0.45\textwidth]{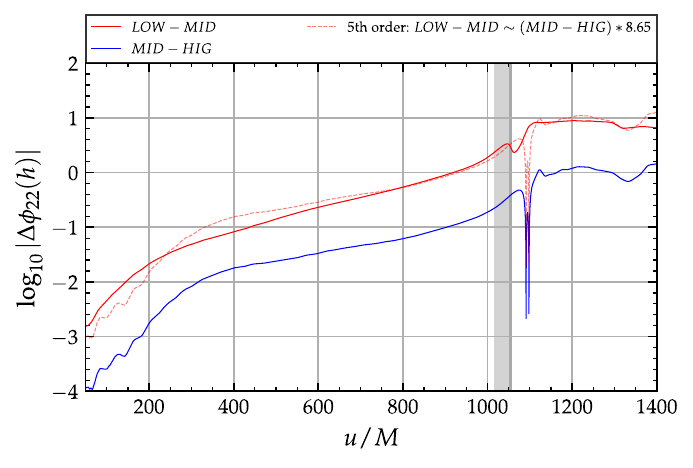}
  \end{tabular}
  \caption{GW phase difference convergence rate study for the MPA1q1.6 simulation with WENOZ reconstruction. 
  \label{fig:gw_phase_errors_MPA1}}
\end{figure}

For a systematic analysis of the convergence properties of MPA1q1.6 we perform three simulations with increasing resolution and adopt the notation $(n_i) = (64, 96, 128) \equiv (\textrm{LOW, MID, HIG})$ of the previous section. \autoref{fig:gw_phase_errors_MPA1} presents a self-convergence study for the waveform phase differences from the MPA1q1.6 simulation. The phase differences decrease with increasing resolution, demonstrating that the numerical scheme used is converging consistently to a solution. This trend is corroborated by the observation that the differences in merger times also reduce (shrinking of gray shaded area). 

The convergence analysis reveals a distinct fifth-order convergence behaviour across the resolution hierarchy. The use of the EFL method is responsible for this high convergence rate and the small absolute difference in the MID-HIG phase near the merger. Furthermore, during the early post-merger regime, the convergence rate continues to exhibit a fifth-order trend, underscoring the robustness of the EFL method. This demonstrates its ability to accurately capture the intense and highly dynamic interactions characterizing the remnant NS. Such a robust performance highlights the suitability of the EFL method for detailed post-merger investigations.

\begin{figure}[h]
  \centering
    \includegraphics[width=0.45\textwidth]{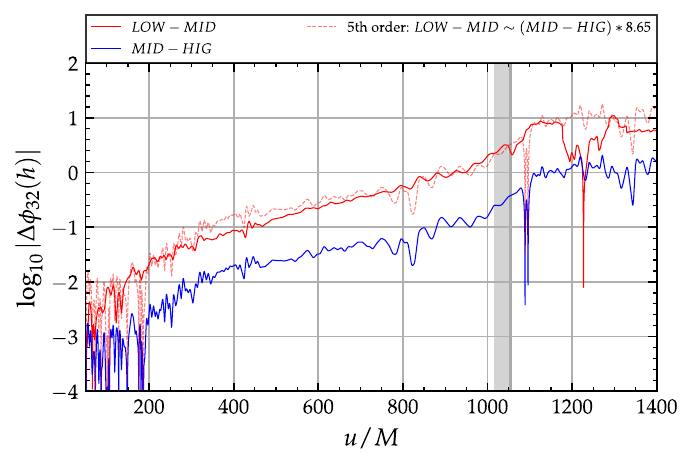}
    \includegraphics[width=0.45\textwidth]{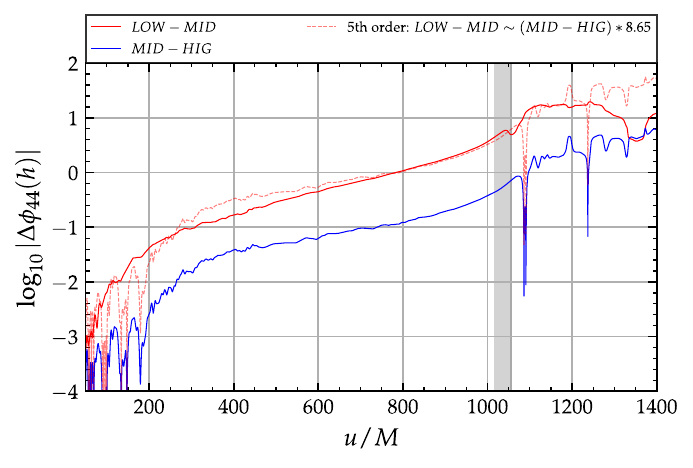}
  \caption{GW phase difference convergence rate study for higher modes of the MPA1q1.6 simulation with WENOZ reconstruction. Top: $(3,2)$-mode. Bottom: $(4,4)$-mode.
  \label{fig:gw_phase_errors_MPA1_higher_modes}}
\end{figure}

Encouraged by the clear fifth-order convergence pattern for the dominant $(2,2)$-mode, we extend the study to the convergence of subdominant modes of the gravitational wave multipoles. \autoref{fig:gw_phase_errors_MPA1_higher_modes} illustrates the convergence study for the $(3,2)$- and $(4,4)$-mode. In these cases, the phase differences consistently diminish with increasing resolution, maintaining a clear fifth-order convergence trend. Moreover, this behaviour persists through the merger and extends into the early post-merger phase, further attesting to the reliability and precision of the EFL method across multiple wave components.

\section{Conclusions}
\label{sec:conclusions}

The primary aim of the present work is to extend the use of the EFL method to the eccentricity reduction of BNS initial data sets. Additionally, we investigate whether the use of the EFL scheme in the eccentricity reduction of BNS initial data improves the convergence properties of the resulting gravitational waveforms. Our findings strongly support this hypothesis. Specifically, the entropy-based method is applied to a couple of BNS merger simulations, showcasing its efficacy.  The new feature that we add with the present work to the BNS simulations carried out in \cite{Doulis:2022vkx} is that the EFL method is used in the construction of the SGRID initial data through the process of eccentricity reduction. 

In \autoref{sec:ID}, we explain in detail how to employ the EFL method in the construction of BNS initial data via the eccentricity reduction process in the SGRID framework. Two eccentricity reduction methods are considered: the first modifies the radial velocity and eccentricity (\autoref{sec:BNSdata}), while the second adjusts the radial velocity and orbital angular velocity (\autoref{sec:DNSdata}). For each approach, we demonstrate the use of the eccentricity reduction process using example BNS configurations. 

\autoref{fig:hybrid_plots_bam95} and \autoref{fig:hybrid_plots_MPA1} demonstrate the robustness of the EFL method in capturing non-smooth features throughout the inspiral, merger, and post-merger phases. The entropy limiter accurately tracks the surface of the inspiraling neutron stars (NSs) and converges to zero in regions of smooth flow. It effectively identifies collisional shocks during the merger and captures the outward propagation of spiral density waves in the post-merger phase, demonstrating its reliability even under complex dynamical conditions. The MPA1q1.6 simulation showcases more complex and asymmetric features, attributable to its unequal mass and spinning nature, while \BAM{95} maintains a symmetric and relatively simpler profile reflective of an equal mass, non-spinning configuration. These differences highlight the ability of the EFL method to adapt and accurately resolve the unique dynamics of each scenario.

A convergence analysis of the gravitational waveforms generated from these simulations confirms that using the EFL method in both the eccentricity reduction of the initial data and the temporal evolution leads to fifth-order convergent waveforms at current production resolutions. Detailed results are discussed in \autoref{sec:gw_analysis_bam95} and \autoref{sec:gw_analysis_MPA1}. This fifth-order convergence is observed not only in the dominant $(2,2)$-mode of the strain but also in subdominant modes such as $(3,2)$ and $(4,4)$. To our knowledge, these results represent the first demonstration of fifth-order convergence in BNS waveform production. The estimated phase error in the waveforms is approximately a factor ${\sim}2$ smaller than the error in the state-of-the-art high-order WENOZ scheme used in the same BAM code, at the same resolution. In conclusion, our study establishes that the incorporation of the EFL scheme into the eccentricity reduction of initial data significantly enhances the accuracy and convergence properties of BNS simulations. This makes the method a valuable tool for producing high-quality waveforms, suitable for precision gravitational wave astronomy and further advancing the field of numerical relativity.

\begin{acknowledgements}

We would like to thank members of the Jena group for fruitful discussions and invaluable input. 
GD acknowledges funding from the European High Performance Computing Joint Undertaking (JU) and Belgium, Czech Republic, France, Germany, Greece, Italy, Norway, and Spain under grant agreement No 101093441 (SPACE).  
SB acknowledges funding from the EU Horizon under
ERC Starting Grant, grant agreement no.~BinGraSp-714626, and 
ERC Consolidator Grant, grant agreement no.~InspiReM-101043372.
WT acknowledges support by the National Science Foundation under grants
PHY-2136036 and PHY-2408903.

Computations where performed on the ARA and DRACO clusters at
Friedrich Schiller University Jena. The ARA cluster is partially
funded by the DFG grants INST 275/334-1 FUGG and INST 275/363-1 FUGG,
and ERC Starting Grant, grant agreement no.~BinGraSp-714626.
The authors also gratefully acknowledge the Gauss Centre for Supercomputing
e.V. (\url{www.gauss-centre.eu}) for funding this project by providing
computing time on the GCS Supercomputer SuperMUC-NG at Leibniz
Supercomputing Centre (\url{www.lrz.de}) with the allocations
{\tt pn36ge} and {\tt pn36jo}.  

\end{acknowledgements}


%


\begin{thebibliography}{33}%
\makeatletter
\providecommand \@ifxundefined [1]{%
 \@ifx{#1\undefined}
}%
\providecommand \@ifnum [1]{%
 \ifnum #1\expandafter \@firstoftwo
 \else \expandafter \@secondoftwo
 \fi
}%
\providecommand \@ifx [1]{%
 \ifx #1\expandafter \@firstoftwo
 \else \expandafter \@secondoftwo
 \fi
}%
\providecommand \natexlab [1]{#1}%
\providecommand \enquote  [1]{``#1''}%
\providecommand \bibnamefont  [1]{#1}%
\providecommand \bibfnamefont [1]{#1}%
\providecommand \citenamefont [1]{#1}%
\providecommand \href@noop [0]{\@secondoftwo}%
\providecommand \href [0]{\begingroup \@sanitize@url \@href}%
\providecommand \@href[1]{\@@startlink{#1}\@@href}%
\providecommand \@@href[1]{\endgroup#1\@@endlink}%
\providecommand \@sanitize@url [0]{\catcode `\\12\catcode `\$12\catcode
  `\&12\catcode `\#12\catcode `\^12\catcode `\_12\catcode `\%12\relax}%
\providecommand \@@startlink[1]{}%
\providecommand \@@endlink[0]{}%
\providecommand \url  [0]{\begingroup\@sanitize@url \@url }%
\providecommand \@url [1]{\endgroup\@href {#1}{\urlprefix }}%
\providecommand \urlprefix  [0]{URL }%
\providecommand \Eprint [0]{\href }%
\providecommand \doibase [0]{https://doi.org/}%
\providecommand \selectlanguage [0]{\@gobble}%
\providecommand \bibinfo  [0]{\@secondoftwo}%
\providecommand \bibfield  [0]{\@secondoftwo}%
\providecommand \translation [1]{[#1]}%
\providecommand \BibitemOpen [0]{}%
\providecommand \bibitemStop [0]{}%
\providecommand \bibitemNoStop [0]{.\EOS\space}%
\providecommand \EOS [0]{\spacefactor3000\relax}%
\providecommand \BibitemShut  [1]{\csname bibitem#1\endcsname}%
\let\auto@bib@innerbib\@empty
\bibitem [{\citenamefont {Abbott}\ \emph
  {et~al.}(2017{\natexlab{a}})\citenamefont {Abbott} \emph
  {et~al.}}]{TheLIGOScientific:2017qsa}%
  \BibitemOpen
  \bibfield  {author} {\bibinfo {author} {\bibfnamefont {B.~P.}\ \bibnamefont
  {Abbott}} \emph {et~al.} (\bibinfo {collaboration} {Virgo, LIGO
  Scientific}),\ }\bibfield  {title} {\bibinfo {title} {{GW170817: Observation
  of Gravitational Waves from a Binary Neutron Star Inspiral}},\ }\href
  {https://doi.org/10.1103/PhysRevLett.119.161101} {\bibfield  {journal}
  {\bibinfo  {journal} {Phys. Rev. Lett.}\ }\textbf {\bibinfo {volume} {119}},\
  \bibinfo {pages} {161101} (\bibinfo {year} {2017}{\natexlab{a}})},\ \Eprint
  {https://arxiv.org/abs/1710.05832} {arXiv:1710.05832 [gr-qc]} \BibitemShut
  {NoStop}%
\bibitem [{\citenamefont {Abbott}\ \emph
  {et~al.}(2017{\natexlab{b}})\citenamefont {Abbott} \emph
  {et~al.}}]{GBM:2017lvd}%
  \BibitemOpen
  \bibfield  {author} {\bibinfo {author} {\bibfnamefont {B.~P.}\ \bibnamefont
  {Abbott}} \emph {et~al.} (\bibinfo {collaboration} {GROND, SALT Group,
  OzGrav, DFN, INTEGRAL, Virgo, Insight-Hxmt, MAXI Team, Fermi-LAT, J-GEM,
  RATIR, IceCube, CAASTRO, LWA, ePESSTO, GRAWITA, RIMAS, SKA South
  Africa/MeerKAT, H.E.S.S., 1M2H Team, IKI-GW Follow-up, Fermi GBM, Pi of Sky,
  DWF (Deeper Wider Faster Program), Dark Energy Survey, MASTER, AstroSat
  Cadmium Zinc Telluride Imager Team, Swift, Pierre Auger, ASKAP, VINROUGE,
  JAGWAR, Chandra Team at McGill University, TTU-NRAO, GROWTH, AGILE Team, MWA,
  ATCA, AST3, TOROS, Pan-STARRS, NuSTAR, ATLAS Telescopes, BOOTES, CaltechNRAO,
  LIGO Scientific, High Time Resolution Universe Survey, Nordic Optical
  Telescope, Las Cumbres Observatory Group, TZAC Consortium, LOFAR, IPN, DLT40,
  Texas Tech University, HAWC, ANTARES, KU, Dark Energy Camera GW-EM, CALET,
  Euro VLBI Team, ALMA}),\ }\bibfield  {title} {\bibinfo {title}
  {{Multi-messenger Observations of a Binary Neutron Star Merger}},\ }\href
  {https://doi.org/10.3847/2041-8213/aa91c9} {\bibfield  {journal} {\bibinfo
  {journal} {Astrophys. J.}\ }\textbf {\bibinfo {volume} {848}},\ \bibinfo
  {pages} {L12} (\bibinfo {year} {2017}{\natexlab{b}})},\ \Eprint
  {https://arxiv.org/abs/1710.05833} {arXiv:1710.05833 [astro-ph.HE]}
  \BibitemShut {NoStop}%
\bibitem [{\citenamefont {{Toro}}(1999)}]{Toro:1999}%
  \BibitemOpen
  \bibfield  {author} {\bibinfo {author} {\bibfnamefont {E.~F.}\ \bibnamefont
  {{Toro}}},\ }\href@noop {} {\emph {\bibinfo {title} {Riemann Solvers and
  Numerical Methods for Fluid Dynamics}}},\ \bibinfo {edition} {2nd}\ ed.\
  (\bibinfo  {publisher} {Springer-Verlag},\ \bibinfo {year}
  {1999})\BibitemShut {NoStop}%
\bibitem [{\citenamefont {Bernuzzi}\ \emph {et~al.}(2012)\citenamefont
  {Bernuzzi}, \citenamefont {Nagar}, \citenamefont {Thierfelder},\ and\
  \citenamefont {Br{\"u}gmann}}]{Bernuzzi:2012ci}%
  \BibitemOpen
  \bibfield  {author} {\bibinfo {author} {\bibfnamefont {S.}~\bibnamefont
  {Bernuzzi}}, \bibinfo {author} {\bibfnamefont {A.}~\bibnamefont {Nagar}},
  \bibinfo {author} {\bibfnamefont {M.}~\bibnamefont {Thierfelder}},\ and\
  \bibinfo {author} {\bibfnamefont {B.}~\bibnamefont {Br{\"u}gmann}},\
  }\bibfield  {title} {\bibinfo {title} {{Tidal effects in binary neutron star
  coalescence}},\ }\href {https://doi.org/10.1103/PhysRevD.86.044030}
  {\bibfield  {journal} {\bibinfo  {journal} {Phys.Rev.}\ }\textbf {\bibinfo
  {volume} {D86}},\ \bibinfo {pages} {044030} (\bibinfo {year} {2012})},\
  \Eprint {https://arxiv.org/abs/1205.3403} {arXiv:1205.3403 [gr-qc]}
  \BibitemShut {NoStop}%
\bibitem [{\citenamefont {Radice}\ \emph
  {et~al.}(2014{\natexlab{a}})\citenamefont {Radice}, \citenamefont
  {Rezzolla},\ and\ \citenamefont {Galeazzi}}]{Radice:2013hxh}%
  \BibitemOpen
  \bibfield  {author} {\bibinfo {author} {\bibfnamefont {D.}~\bibnamefont
  {Radice}}, \bibinfo {author} {\bibfnamefont {L.}~\bibnamefont {Rezzolla}},\
  and\ \bibinfo {author} {\bibfnamefont {F.}~\bibnamefont {Galeazzi}},\
  }\bibfield  {title} {\bibinfo {title} {{Beyond second-order convergence in
  simulations of binary neutron stars in full general-relativity}},\ }\href
  {https://doi.org/10.1093/mnrasl/slt137} {\bibfield  {journal} {\bibinfo
  {journal} {Mon.Not.Roy.Astron.Soc.}\ }\textbf {\bibinfo {volume} {437}},\
  \bibinfo {pages} {L46} (\bibinfo {year} {2014}{\natexlab{a}})},\ \Eprint
  {https://arxiv.org/abs/1306.6052} {arXiv:1306.6052 [gr-qc]} \BibitemShut
  {NoStop}%
\bibitem [{\citenamefont {Radice}\ \emph
  {et~al.}(2014{\natexlab{b}})\citenamefont {Radice}, \citenamefont
  {Rezzolla},\ and\ \citenamefont {Galeazzi}}]{Radice:2013xpa}%
  \BibitemOpen
  \bibfield  {author} {\bibinfo {author} {\bibfnamefont {D.}~\bibnamefont
  {Radice}}, \bibinfo {author} {\bibfnamefont {L.}~\bibnamefont {Rezzolla}},\
  and\ \bibinfo {author} {\bibfnamefont {F.}~\bibnamefont {Galeazzi}},\
  }\bibfield  {title} {\bibinfo {title} {{High-Order Fully General-Relativistic
  Hydrodynamics: new Approaches and Tests}},\ }\href
  {https://doi.org/10.1088/0264-9381/31/7/075012} {\bibfield  {journal}
  {\bibinfo  {journal} {Class.Quant.Grav.}\ }\textbf {\bibinfo {volume} {31}},\
  \bibinfo {pages} {075012} (\bibinfo {year} {2014}{\natexlab{b}})},\ \Eprint
  {https://arxiv.org/abs/1312.5004} {arXiv:1312.5004 [gr-qc]} \BibitemShut
  {NoStop}%
\bibitem [{\citenamefont {Bernuzzi}\ and\ \citenamefont
  {Dietrich}(2016)}]{Bernuzzi:2016pie}%
  \BibitemOpen
  \bibfield  {author} {\bibinfo {author} {\bibfnamefont {S.}~\bibnamefont
  {Bernuzzi}}\ and\ \bibinfo {author} {\bibfnamefont {T.}~\bibnamefont
  {Dietrich}},\ }\bibfield  {title} {\bibinfo {title} {{Gravitational waveforms
  from binary neutron star mergers with high-order
  weighted-essentially-nonoscillatory schemes in numerical relativity}},\
  }\href {https://doi.org/10.1103/PhysRevD.94.064062} {\bibfield  {journal}
  {\bibinfo  {journal} {Phys. Rev.}\ }\textbf {\bibinfo {volume} {D94}},\
  \bibinfo {pages} {064062} (\bibinfo {year} {2016})},\ \Eprint
  {https://arxiv.org/abs/1604.07999} {arXiv:1604.07999 [gr-qc]} \BibitemShut
  {NoStop}%
\bibitem [{\citenamefont {{Sweby}}(1984)}]{Sweby:1984a}%
  \BibitemOpen
  \bibfield  {author} {\bibinfo {author} {\bibfnamefont {P.~K.}\ \bibnamefont
  {{Sweby}}},\ }\bibfield  {title} {\bibinfo {title} {{High Resolution Schemes
  Using Flux Limiters for Hyperbolic Conservation Laws}},\ }\href
  {https://doi.org/10.1137/0721062} {\bibfield  {journal} {\bibinfo  {journal}
  {SIAM Journal on Numerical Analysis}\ }\textbf {\bibinfo {volume} {21}},\
  \bibinfo {pages} {995} (\bibinfo {year} {1984})}\BibitemShut {NoStop}%
\bibitem [{\citenamefont {Doulis}\ \emph {et~al.}(2022)\citenamefont {Doulis},
  \citenamefont {Atteneder}, \citenamefont {Bernuzzi},\ and\ \citenamefont
  {Br\"ugmann}}]{Doulis:2022vkx}%
  \BibitemOpen
  \bibfield  {author} {\bibinfo {author} {\bibfnamefont {G.}~\bibnamefont
  {Doulis}}, \bibinfo {author} {\bibfnamefont {F.}~\bibnamefont {Atteneder}},
  \bibinfo {author} {\bibfnamefont {S.}~\bibnamefont {Bernuzzi}},\ and\
  \bibinfo {author} {\bibfnamefont {B.}~\bibnamefont {Br\"ugmann}},\ }\bibfield
   {title} {\bibinfo {title} {{Entropy-limited higher-order central scheme for
  neutron star merger simulations}},\ }\href
  {https://doi.org/10.1103/PhysRevD.106.024001} {\bibfield  {journal} {\bibinfo
   {journal} {Phys. Rev. D}\ }\textbf {\bibinfo {volume} {106}},\ \bibinfo
  {pages} {024001} (\bibinfo {year} {2022})},\ \Eprint
  {https://arxiv.org/abs/2202.08839} {arXiv:2202.08839 [gr-qc]} \BibitemShut
  {NoStop}%
\bibitem [{\citenamefont {Guermond}\ and\ \citenamefont
  {Pasquetti}(2008)}]{Guermond:2008}%
  \BibitemOpen
  \bibfield  {author} {\bibinfo {author} {\bibfnamefont {J.-L.}\ \bibnamefont
  {Guermond}}\ and\ \bibinfo {author} {\bibfnamefont {R.}~\bibnamefont
  {Pasquetti}},\ }\bibfield  {title} {\bibinfo {title} {Entropy-based nonlinear
  viscosity for fourier approximations of conservation laws},\ }\href
  {https://doi.org/https://doi.org/10.1016/j.crma.2008.05.013} {\bibfield
  {journal} {\bibinfo  {journal} {Comptes Rendus Mathematique}\ }\textbf
  {\bibinfo {volume} {346}},\ \bibinfo {pages} {801 } (\bibinfo {year}
  {2008})}\BibitemShut {NoStop}%
\bibitem [{\citenamefont {Guermond}\ \emph {et~al.}(2011)\citenamefont
  {Guermond}, \citenamefont {Pasquetti},\ and\ \citenamefont
  {Popov}}]{Guermond:2011}%
  \BibitemOpen
  \bibfield  {author} {\bibinfo {author} {\bibfnamefont {J.-L.}\ \bibnamefont
  {Guermond}}, \bibinfo {author} {\bibfnamefont {R.}~\bibnamefont
  {Pasquetti}},\ and\ \bibinfo {author} {\bibfnamefont {B.}~\bibnamefont
  {Popov}},\ }\bibfield  {title} {\bibinfo {title} {Entropy viscosity method
  for nonlinear conservation laws},\ }\href
  {https://doi.org/http://dx.doi.org/10.1016/j.jcp.2010.11.043} {\bibfield
  {journal} {\bibinfo  {journal} {Journal of Computational Physics}\ }\textbf
  {\bibinfo {volume} {230}},\ \bibinfo {pages} {4248 } (\bibinfo {year}
  {2011})},\ \bibinfo {note} {special issue High Order Methods for \{CFD\}
  Problems}\BibitemShut {NoStop}%
\bibitem [{\citenamefont {Guercilena}\ \emph {et~al.}(2017)\citenamefont
  {Guercilena}, \citenamefont {Radice},\ and\ \citenamefont
  {Rezzolla}}]{Guercilena:2016fdl}%
  \BibitemOpen
  \bibfield  {author} {\bibinfo {author} {\bibfnamefont {F.}~\bibnamefont
  {Guercilena}}, \bibinfo {author} {\bibfnamefont {D.}~\bibnamefont {Radice}},\
  and\ \bibinfo {author} {\bibfnamefont {L.}~\bibnamefont {Rezzolla}},\
  }\bibfield  {title} {\bibinfo {title} {{Entropy-limited hydrodynamics: a
  novel approach to relativistic hydrodynamics}},\ }\href
  {https://doi.org/10.1186/s40668-017-0022-0} {\bibfield  {journal} {\bibinfo
  {journal} {Comput. Astrophys. Cosmol.}\ }\textbf {\bibinfo {volume} {4}},\
  \bibinfo {pages} {3} (\bibinfo {year} {2017})},\ \Eprint
  {https://arxiv.org/abs/1612.06251} {arXiv:1612.06251 [gr-qc]} \BibitemShut
  {NoStop}%
\bibitem [{\citenamefont {Gourgoulhon}\ \emph {et~al.}(2001)\citenamefont
  {Gourgoulhon}, \citenamefont {Grandclement}, \citenamefont {Taniguchi},
  \citenamefont {Marck},\ and\ \citenamefont {Bonazzola}}]{Gourgoulhon:2000nn}%
  \BibitemOpen
  \bibfield  {author} {\bibinfo {author} {\bibfnamefont {E.}~\bibnamefont
  {Gourgoulhon}}, \bibinfo {author} {\bibfnamefont {P.}~\bibnamefont
  {Grandclement}}, \bibinfo {author} {\bibfnamefont {K.}~\bibnamefont
  {Taniguchi}}, \bibinfo {author} {\bibfnamefont {J.-A.}\ \bibnamefont
  {Marck}},\ and\ \bibinfo {author} {\bibfnamefont {S.}~\bibnamefont
  {Bonazzola}},\ }\bibfield  {title} {\bibinfo {title} {{Quasiequilibrium
  sequences of synchronized and irrotational binary neutron stars in general
  relativity: 1. Method and tests}},\ }\href
  {https://doi.org/10.1103/PhysRevD.63.064029} {\bibfield  {journal} {\bibinfo
  {journal} {Phys.Rev.}\ }\textbf {\bibinfo {volume} {D63}},\ \bibinfo {pages}
  {064029} (\bibinfo {year} {2001})},\ \Eprint
  {https://arxiv.org/abs/gr-qc/0007028} {arXiv:gr-qc/0007028 [gr-qc]}
  \BibitemShut {NoStop}%
\bibitem [{\citenamefont {Tichy}(2012)}]{Tichy:2012rp}%
  \BibitemOpen
  \bibfield  {author} {\bibinfo {author} {\bibfnamefont {W.}~\bibnamefont
  {Tichy}},\ }\bibfield  {title} {\bibinfo {title} {{Constructing
  quasi-equilibrium initial data for binary neutron stars with arbitrary
  spins}},\ }\href {https://doi.org/10.1103/PhysRevD.86.064024} {\bibfield
  {journal} {\bibinfo  {journal} {Phys. Rev. D}\ }\textbf {\bibinfo {volume}
  {86}},\ \bibinfo {pages} {064024} (\bibinfo {year} {2012})},\ \Eprint
  {https://arxiv.org/abs/1209.5336} {arXiv:1209.5336 [gr-qc]} \BibitemShut
  {NoStop}%
\bibitem [{\citenamefont {Banyuls}\ \emph {et~al.}(1997)\citenamefont
  {Banyuls}, \citenamefont {Font}, \citenamefont {Ibanez}, \citenamefont
  {Marti},\ and\ \citenamefont {Miralles}}]{Banyuls:1997zz}%
  \BibitemOpen
  \bibfield  {author} {\bibinfo {author} {\bibfnamefont {F.}~\bibnamefont
  {Banyuls}}, \bibinfo {author} {\bibfnamefont {J.~A.}\ \bibnamefont {Font}},
  \bibinfo {author} {\bibfnamefont {J.~M.~A.}\ \bibnamefont {Ibanez}}, \bibinfo
  {author} {\bibfnamefont {J.~M.~A.}\ \bibnamefont {Marti}},\ and\ \bibinfo
  {author} {\bibfnamefont {J.~A.}\ \bibnamefont {Miralles}},\ }\bibfield
  {title} {\bibinfo {title} {{Numerical {3+1} General Relativistic
  Hydrodynamics: A Local Characteristic Approach}},\ }\href@noop {} {\bibfield
  {journal} {\bibinfo  {journal} {Astrophys. J.}\ }\textbf {\bibinfo {volume}
  {476}},\ \bibinfo {pages} {221} (\bibinfo {year} {1997})}\BibitemShut
  {NoStop}%
\bibitem [{\citenamefont {Doulis}\ \emph {et~al.}(2024)\citenamefont {Doulis},
  \citenamefont {Bernuzzi},\ and\ \citenamefont {Tichy}}]{Doulis:2024aew}%
  \BibitemOpen
  \bibfield  {author} {\bibinfo {author} {\bibfnamefont {G.}~\bibnamefont
  {Doulis}}, \bibinfo {author} {\bibfnamefont {S.}~\bibnamefont {Bernuzzi}},\
  and\ \bibinfo {author} {\bibfnamefont {W.}~\bibnamefont {Tichy}},\ }\bibfield
   {title} {\bibinfo {title} {{Entropy based flux limiting scheme for
  conservation laws}},\ }\href@noop {} {\  (\bibinfo {year} {2024})},\ \Eprint
  {https://arxiv.org/abs/2401.04770} {arXiv:2401.04770 [gr-qc]} \BibitemShut
  {NoStop}%
\bibitem [{\citenamefont {Mignone}\ \emph {et~al.}(2010)\citenamefont
  {Mignone}, \citenamefont {Tzeferacos},\ and\ \citenamefont
  {Bodo}}]{Mignone:2010br}%
  \BibitemOpen
  \bibfield  {author} {\bibinfo {author} {\bibfnamefont {A.}~\bibnamefont
  {Mignone}}, \bibinfo {author} {\bibfnamefont {P.}~\bibnamefont
  {Tzeferacos}},\ and\ \bibinfo {author} {\bibfnamefont {G.}~\bibnamefont
  {Bodo}},\ }\bibfield  {title} {\bibinfo {title} {{High-order conservative
  finite difference GLM-MHD schemes for cell-centered MHD}},\ }\href
  {https://doi.org/10.1016/j.jcp.2010.04.013} {\bibfield  {journal} {\bibinfo
  {journal} {J.Comput.Phys.}\ }\textbf {\bibinfo {volume} {229}},\ \bibinfo
  {pages} {5896} (\bibinfo {year} {2010})},\ \Eprint
  {https://arxiv.org/abs/1001.2832} {arXiv:1001.2832 [astro-ph.HE]}
  \BibitemShut {NoStop}%
\bibitem [{\citenamefont {Thierfelder}\ \emph {et~al.}(2011)\citenamefont
  {Thierfelder}, \citenamefont {Bernuzzi},\ and\ \citenamefont
  {Br{\"u}gmann}}]{Thierfelder:2011yi}%
  \BibitemOpen
  \bibfield  {author} {\bibinfo {author} {\bibfnamefont {M.}~\bibnamefont
  {Thierfelder}}, \bibinfo {author} {\bibfnamefont {S.}~\bibnamefont
  {Bernuzzi}},\ and\ \bibinfo {author} {\bibfnamefont {B.}~\bibnamefont
  {Br{\"u}gmann}},\ }\bibfield  {title} {\bibinfo {title} {{Numerical
  relativity simulations of binary neutron stars}},\ }\href
  {https://doi.org/10.1103/PhysRevD.84.044012} {\bibfield  {journal} {\bibinfo
  {journal} {Phys.Rev.}\ }\textbf {\bibinfo {volume} {D84}},\ \bibinfo {pages}
  {044012} (\bibinfo {year} {2011})},\ \Eprint
  {https://arxiv.org/abs/1104.4751} {arXiv:1104.4751 [gr-qc]} \BibitemShut
  {NoStop}%
\bibitem [{\citenamefont {Borges}\ \emph {et~al.}(2008)\citenamefont {Borges},
  \citenamefont {Carmona}, \citenamefont {Costa},\ and\ \citenamefont
  {Don}}]{Borges:2008a}%
  \BibitemOpen
  \bibfield  {author} {\bibinfo {author} {\bibfnamefont {R.}~\bibnamefont
  {Borges}}, \bibinfo {author} {\bibfnamefont {M.}~\bibnamefont {Carmona}},
  \bibinfo {author} {\bibfnamefont {B.}~\bibnamefont {Costa}},\ and\ \bibinfo
  {author} {\bibfnamefont {W.~S.}\ \bibnamefont {Don}},\ }\bibfield  {title}
  {\bibinfo {title} {An improved weighted essentially non-oscillatory scheme
  for hyperbolic conservation laws},\ }\href
  {https://doi.org/10.1016/j.jcp.2007.11.038} {\bibfield  {journal} {\bibinfo
  {journal} {Journal of Computational Physics}\ }\textbf {\bibinfo {volume}
  {227}},\ \bibinfo {pages} {3191} (\bibinfo {year} {2008})}\BibitemShut
  {NoStop}%
\bibitem [{\citenamefont {Hesthaven}(2018)}]{Hesthaven:2017}%
  \BibitemOpen
  \bibfield  {author} {\bibinfo {author} {\bibfnamefont {J.~S.}\ \bibnamefont
  {Hesthaven}},\ }\href {https://doi.org/10.1137/1.9781611975109} {\emph
  {\bibinfo {title} {Numerical Methods for Conservation Laws}}}\ (\bibinfo
  {publisher} {Society for Industrial and Applied Mathematics},\ \bibinfo
  {address} {Philadelphia, PA},\ \bibinfo {year} {2018})\ \Eprint
  {https://arxiv.org/abs/https://epubs.siam.org/doi/pdf/10.1137/1.9781611975109}
  {https://epubs.siam.org/doi/pdf/10.1137/1.9781611975109} \BibitemShut
  {NoStop}%
\bibitem [{\citenamefont {Br{\"u}gmann}\ \emph {et~al.}(2008)\citenamefont
  {Br{\"u}gmann}, \citenamefont {Gonzalez}, \citenamefont {Hannam},
  \citenamefont {Husa}, \citenamefont {Sperhake} \emph
  {et~al.}}]{Brugmann:2008zz}%
  \BibitemOpen
  \bibfield  {author} {\bibinfo {author} {\bibfnamefont {B.}~\bibnamefont
  {Br{\"u}gmann}}, \bibinfo {author} {\bibfnamefont {J.~A.}\ \bibnamefont
  {Gonzalez}}, \bibinfo {author} {\bibfnamefont {M.}~\bibnamefont {Hannam}},
  \bibinfo {author} {\bibfnamefont {S.}~\bibnamefont {Husa}}, \bibinfo {author}
  {\bibfnamefont {U.}~\bibnamefont {Sperhake}}, \emph {et~al.},\ }\bibfield
  {title} {\bibinfo {title} {{Calibration of Moving Puncture Simulations}},\
  }\href {https://doi.org/10.1103/PhysRevD.77.024027} {\bibfield  {journal}
  {\bibinfo  {journal} {Phys.Rev.}\ }\textbf {\bibinfo {volume} {D77}},\
  \bibinfo {pages} {024027} (\bibinfo {year} {2008})},\ \Eprint
  {https://arxiv.org/abs/gr-qc/0610128} {arXiv:gr-qc/0610128 [gr-qc]}
  \BibitemShut {NoStop}%
\bibitem [{\citenamefont {Dietrich}\ \emph
  {et~al.}(2015{\natexlab{a}})\citenamefont {Dietrich}, \citenamefont
  {Bernuzzi}, \citenamefont {Ujevic},\ and\ \citenamefont
  {Br{\"u}gmann}}]{Dietrich:2015iva}%
  \BibitemOpen
  \bibfield  {author} {\bibinfo {author} {\bibfnamefont {T.}~\bibnamefont
  {Dietrich}}, \bibinfo {author} {\bibfnamefont {S.}~\bibnamefont {Bernuzzi}},
  \bibinfo {author} {\bibfnamefont {M.}~\bibnamefont {Ujevic}},\ and\ \bibinfo
  {author} {\bibfnamefont {B.}~\bibnamefont {Br{\"u}gmann}},\ }\bibfield
  {title} {\bibinfo {title} {{Numerical relativity simulations of neutron star
  merger remnants using conservative mesh refinement}},\ }\href
  {https://doi.org/10.1103/PhysRevD.91.124041} {\bibfield  {journal} {\bibinfo
  {journal} {Phys. Rev.}\ }\textbf {\bibinfo {volume} {D91}},\ \bibinfo {pages}
  {124041} (\bibinfo {year} {2015}{\natexlab{a}})},\ \Eprint
  {https://arxiv.org/abs/1504.01266} {arXiv:1504.01266 [gr-qc]} \BibitemShut
  {NoStop}%
\bibitem [{\citenamefont {Dietrich}\ \emph
  {et~al.}(2015{\natexlab{b}})\citenamefont {Dietrich}, \citenamefont
  {Moldenhauer}, \citenamefont {Johnson-McDaniel}, \citenamefont {Bernuzzi},
  \citenamefont {Markakis}, \citenamefont {Br{\"u}gmann},\ and\ \citenamefont
  {Tichy}}]{Dietrich:2015pxa}%
  \BibitemOpen
  \bibfield  {author} {\bibinfo {author} {\bibfnamefont {T.}~\bibnamefont
  {Dietrich}}, \bibinfo {author} {\bibfnamefont {N.}~\bibnamefont
  {Moldenhauer}}, \bibinfo {author} {\bibfnamefont {N.~K.}\ \bibnamefont
  {Johnson-McDaniel}}, \bibinfo {author} {\bibfnamefont {S.}~\bibnamefont
  {Bernuzzi}}, \bibinfo {author} {\bibfnamefont {C.~M.}\ \bibnamefont
  {Markakis}}, \bibinfo {author} {\bibfnamefont {B.}~\bibnamefont
  {Br{\"u}gmann}},\ and\ \bibinfo {author} {\bibfnamefont {W.}~\bibnamefont
  {Tichy}},\ }\bibfield  {title} {\bibinfo {title} {{Binary Neutron Stars with
  Generic Spin, Eccentricity, Mass ratio, and Compactness - Quasi-equilibrium
  Sequences and First Evolutions}},\ }\href
  {https://doi.org/10.1103/PhysRevD.92.124007} {\bibfield  {journal} {\bibinfo
  {journal} {Phys. Rev.}\ }\textbf {\bibinfo {volume} {D92}},\ \bibinfo {pages}
  {124007} (\bibinfo {year} {2015}{\natexlab{b}})},\ \Eprint
  {https://arxiv.org/abs/1507.07100} {arXiv:1507.07100 [gr-qc]} \BibitemShut
  {NoStop}%
\bibitem [{\citenamefont {Tichy}\ \emph {et~al.}(2019)\citenamefont {Tichy},
  \citenamefont {Rashti}, \citenamefont {Dietrich}, \citenamefont {Dudi},\ and\
  \citenamefont {Brügmann}}]{Tichy:2019ouu}%
  \BibitemOpen
  \bibfield  {author} {\bibinfo {author} {\bibfnamefont {W.}~\bibnamefont
  {Tichy}}, \bibinfo {author} {\bibfnamefont {A.}~\bibnamefont {Rashti}},
  \bibinfo {author} {\bibfnamefont {T.}~\bibnamefont {Dietrich}}, \bibinfo
  {author} {\bibfnamefont {R.}~\bibnamefont {Dudi}},\ and\ \bibinfo {author}
  {\bibfnamefont {B.}~\bibnamefont {Brügmann}},\ }\bibfield  {title} {\bibinfo
  {title} {{Constructing Binary Neutron Star Initial Data with High Spins, High
  Compactness, and High Mass-Ratios}},\ }\href
  {https://doi.org/10.1103/PhysRevD.100.124046} {\bibfield  {journal} {\bibinfo
   {journal} {Phys. Rev.}\ }\textbf {\bibinfo {volume} {D100}},\ \bibinfo
  {pages} {124046} (\bibinfo {year} {2019})},\ \Eprint
  {https://arxiv.org/abs/1910.09690} {arXiv:1910.09690 [gr-qc]} \BibitemShut
  {NoStop}%
\bibitem [{\citenamefont {Tichy}(2006)}]{Tichy:2006qn}%
  \BibitemOpen
  \bibfield  {author} {\bibinfo {author} {\bibfnamefont {W.}~\bibnamefont
  {Tichy}},\ }\bibfield  {title} {\bibinfo {title} {{Black hole evolution with
  the BSSN system by pseudo-spectral methods}},\ }\href
  {https://doi.org/10.1103/PhysRevD.74.084005} {\bibfield  {journal} {\bibinfo
  {journal} {Phys.Rev.}\ }\textbf {\bibinfo {volume} {D74}},\ \bibinfo {pages}
  {084005} (\bibinfo {year} {2006})},\ \Eprint
  {https://arxiv.org/abs/gr-qc/0609087} {arXiv:gr-qc/0609087 [gr-qc]}
  \BibitemShut {NoStop}%
\bibitem [{\citenamefont {Tichy}(2009{\natexlab{a}})}]{Tichy:2009yr}%
  \BibitemOpen
  \bibfield  {author} {\bibinfo {author} {\bibfnamefont {W.}~\bibnamefont
  {Tichy}},\ }\bibfield  {title} {\bibinfo {title} {{A New numerical method to
  construct binary neutron star initial data}},\ }\href
  {https://doi.org/10.1088/0264-9381/26/17/175018} {\bibfield  {journal}
  {\bibinfo  {journal} {Class.Quant.Grav.}\ }\textbf {\bibinfo {volume} {26}},\
  \bibinfo {pages} {175018} (\bibinfo {year} {2009}{\natexlab{a}})},\ \Eprint
  {https://arxiv.org/abs/0908.0620} {arXiv:0908.0620 [gr-qc]} \BibitemShut
  {NoStop}%
\bibitem [{\citenamefont {Tichy}(2009{\natexlab{b}})}]{Tichy:2009zr}%
  \BibitemOpen
  \bibfield  {author} {\bibinfo {author} {\bibfnamefont {W.}~\bibnamefont
  {Tichy}},\ }\bibfield  {title} {\bibinfo {title} {{Long term black hole
  evolution with the BSSN system by pseudo-spectral methods}},\ }\href
  {https://doi.org/10.1103/PhysRevD.80.104034} {\bibfield  {journal} {\bibinfo
  {journal} {Phys.Rev.}\ }\textbf {\bibinfo {volume} {D80}},\ \bibinfo {pages}
  {104034} (\bibinfo {year} {2009}{\natexlab{b}})},\ \Eprint
  {https://arxiv.org/abs/0911.0973} {arXiv:0911.0973 [gr-qc]} \BibitemShut
  {NoStop}%
\bibitem [{\citenamefont {Tichy}(2011)}]{Tichy:2011gw}%
  \BibitemOpen
  \bibfield  {author} {\bibinfo {author} {\bibfnamefont {W.}~\bibnamefont
  {Tichy}},\ }\bibfield  {title} {\bibinfo {title} {{Initial data for binary
  neutron stars with arbitrary spins}},\ }\href
  {https://doi.org/10.1103/PhysRevD.84.024041} {\bibfield  {journal} {\bibinfo
  {journal} {Phys.Rev.}\ }\textbf {\bibinfo {volume} {D84}},\ \bibinfo {pages}
  {024041} (\bibinfo {year} {2011})},\ \Eprint
  {https://arxiv.org/abs/1107.1440} {arXiv:1107.1440 [gr-qc]} \BibitemShut
  {NoStop}%
\bibitem [{\citenamefont {Kyutoku}\ \emph {et~al.}(2014)\citenamefont
  {Kyutoku}, \citenamefont {Shibata},\ and\ \citenamefont
  {Taniguchi}}]{Kyutoku:2014yba}%
  \BibitemOpen
  \bibfield  {author} {\bibinfo {author} {\bibfnamefont {K.}~\bibnamefont
  {Kyutoku}}, \bibinfo {author} {\bibfnamefont {M.}~\bibnamefont {Shibata}},\
  and\ \bibinfo {author} {\bibfnamefont {K.}~\bibnamefont {Taniguchi}},\
  }\bibfield  {title} {\bibinfo {title} {{Reducing orbital eccentricity in
  initial data of binary neutron stars}},\ }\href
  {https://doi.org/10.1103/PhysRevD.90.064006} {\bibfield  {journal} {\bibinfo
  {journal} {Phys. Rev.}\ }\textbf {\bibinfo {volume} {D90}},\ \bibinfo {pages}
  {064006} (\bibinfo {year} {2014})},\ \Eprint
  {https://arxiv.org/abs/1405.6207} {arXiv:1405.6207 [gr-qc]} \BibitemShut
  {NoStop}%
\bibitem [{\citenamefont {Dietrich}\ \emph {et~al.}(2017)\citenamefont
  {Dietrich}, \citenamefont {Bernuzzi},\ and\ \citenamefont
  {Tichy}}]{Dietrich:2017aum}%
  \BibitemOpen
  \bibfield  {author} {\bibinfo {author} {\bibfnamefont {T.}~\bibnamefont
  {Dietrich}}, \bibinfo {author} {\bibfnamefont {S.}~\bibnamefont {Bernuzzi}},\
  and\ \bibinfo {author} {\bibfnamefont {W.}~\bibnamefont {Tichy}},\ }\bibfield
   {title} {\bibinfo {title} {{Closed-form tidal approximants for binary
  neutron star gravitational waveforms constructed from high-resolution
  numerical relativity simulations}},\ }\href
  {https://doi.org/10.1103/PhysRevD.96.121501} {\bibfield  {journal} {\bibinfo
  {journal} {Phys. Rev.}\ }\textbf {\bibinfo {volume} {D96}},\ \bibinfo {pages}
  {121501} (\bibinfo {year} {2017})},\ \Eprint
  {https://arxiv.org/abs/1706.02969} {arXiv:1706.02969 [gr-qc]} \BibitemShut
  {NoStop}%
\bibitem [{\citenamefont {Dietrich}\ \emph {et~al.}(2018)\citenamefont
  {Dietrich}, \citenamefont {Radice}, \citenamefont {Bernuzzi}, \citenamefont
  {Zappa}, \citenamefont {Perego}, \citenamefont {Brügmann}, \citenamefont
  {Chaurasia}, \citenamefont {Dudi}, \citenamefont {Tichy},\ and\ \citenamefont
  {Ujevic}}]{Dietrich:2018phi}%
  \BibitemOpen
  \bibfield  {author} {\bibinfo {author} {\bibfnamefont {T.}~\bibnamefont
  {Dietrich}}, \bibinfo {author} {\bibfnamefont {D.}~\bibnamefont {Radice}},
  \bibinfo {author} {\bibfnamefont {S.}~\bibnamefont {Bernuzzi}}, \bibinfo
  {author} {\bibfnamefont {F.}~\bibnamefont {Zappa}}, \bibinfo {author}
  {\bibfnamefont {A.}~\bibnamefont {Perego}}, \bibinfo {author} {\bibfnamefont
  {B.}~\bibnamefont {Brügmann}}, \bibinfo {author} {\bibfnamefont {S.~V.}\
  \bibnamefont {Chaurasia}}, \bibinfo {author} {\bibfnamefont {R.}~\bibnamefont
  {Dudi}}, \bibinfo {author} {\bibfnamefont {W.}~\bibnamefont {Tichy}},\ and\
  \bibinfo {author} {\bibfnamefont {M.}~\bibnamefont {Ujevic}},\ }\bibfield
  {title} {\bibinfo {title} {{CoRe database of binary neutron star merger
  waveforms}},\ }\href {https://doi.org/10.1088/1361-6382/aaebc0} {\bibfield
  {journal} {\bibinfo  {journal} {Class. Quant. Grav.}\ }\textbf {\bibinfo
  {volume} {35}},\ \bibinfo {pages} {24LT01} (\bibinfo {year} {2018})},\
  \Eprint {https://arxiv.org/abs/1806.01625} {arXiv:1806.01625 [gr-qc]}
  \BibitemShut {NoStop}%
\bibitem [{\citenamefont {Reisswig}\ and\ \citenamefont
  {Pollney}(2011)}]{Reisswig:2010di}%
  \BibitemOpen
  \bibfield  {author} {\bibinfo {author} {\bibfnamefont {C.}~\bibnamefont
  {Reisswig}}\ and\ \bibinfo {author} {\bibfnamefont {D.}~\bibnamefont
  {Pollney}},\ }\bibfield  {title} {\bibinfo {title} {{Notes on the integration
  of numerical relativity waveforms}},\ }\href
  {https://doi.org/10.1088/0264-9381/28/19/195015} {\bibfield  {journal}
  {\bibinfo  {journal} {Class.Quant.Grav.}\ }\textbf {\bibinfo {volume} {28}},\
  \bibinfo {pages} {195015} (\bibinfo {year} {2011})},\ \Eprint
  {https://arxiv.org/abs/1006.1632} {arXiv:1006.1632 [gr-qc]} \BibitemShut
  {NoStop}%
\bibitem [{\citenamefont {Baumgarte}\ and\ \citenamefont
  {Shapiro}(2010)}]{Baumgarte:2010}%
  \BibitemOpen
  \bibfield  {author} {\bibinfo {author} {\bibfnamefont {T.}~\bibnamefont
  {Baumgarte}}\ and\ \bibinfo {author} {\bibfnamefont {S.}~\bibnamefont
  {Shapiro}},\ }\href@noop {} {\emph {\bibinfo {title} {Numerical
  Relativity}}}\ (\bibinfo  {publisher} {Cambridge University Press},\ \bibinfo
  {address} {Cambridge},\ \bibinfo {year} {2010})\BibitemShut {NoStop}%
\end{thebibliography}
\end{document}